\documentclass{aa}
\usepackage[varg]{txfonts}
\usepackage[utf8]{inputenc}
\usepackage{comment}
\usepackage{amsmath}
\usepackage{upgreek}
\usepackage[colorlinks=true,linkcolor=blue,citecolor=blue]{hyperref}
\usepackage{graphicx}
\usepackage{gensymb}
\usepackage{xspace}
\newcommand{\hei}{\ion{He}{i}\xspace}

\newcommand{\caii}{\ion{Ca}{ii}\xspace}
\newcommand{\rbi}{\ion{Rb}{i}\xspace}
\newcommand{\sriii}{\ion{Sr}{iii}\xspace}
\newcommand{\srii}{\ion{Sr}{ii}\xspace}
\newcommand{\yii}{\ion{Y}{ii}\xspace}
\newcommand{\zrii}{\ion{Zr}{ii}\xspace}
\newcommand{\teiii}{\ion{Te}{iii}\xspace}
\newcommand{\teiiinospace}{\ion{Te}{iii}}
\newcommand{\Laiii}{\ion{La}{iii}\xspace}
\newcommand{\ceii}{\ion{Ce}{ii}\xspace}
\newcommand{\Ceiii}{\ion{Ce}{iii}\xspace}
\newcommand{\rprocess}{\textit{r}-process\xspace}
\newcommand{\micron}{$\upmu$m\xspace}
\usepackage{tabularx}
\usepackage{orcidlink}
\makeatletter
\newcommand\footnoteref[1]{\protected@xdef\@thefnmark{\ref{#1}}\@footnotemark}
\makeatother

\usepackage{natbib}
\bibpunct{(}{)}{;}{a}{}{,} 

\begin{document}



\title{Emergence hour-by-hour of \rprocess features in the kilonova AT2017gfo} 





\titlerunning{AT2017gfo spectral series}
\authorrunning{Sneppen et al.}

\author{
Albert Sneppen\inst{\ref{addr:DAWN},\ref{addr:jagtvej}}\orcidlink{0000-0002-5460-6126},
Darach Watson \inst{\ref{addr:DAWN},\ref{addr:jagtvej}}\orcidlink{0000-0002-4465-8264},
Rasmus Damgaard \inst{\ref{addr:DAWN},\ref{addr:jagtvej}}\orcidlink{0009-0002-5765-4601}, 
Kasper E. Heintz \inst{\ref{addr:DAWN},\ref{addr:jagtvej}}\orcidlink{0000-0002-9389-7413}, 
Nicholas Vieira \inst{\ref{addr:mcgill}}\orcidlink{0000-0001-7815-7604}, 
Petri V\"ais\"anen \inst{\ref{addr:saao},\ref{addr:finca}}\orcidlink{0000-0001-7673-4850} and 
Antoine Mahoro \inst{\ref{addr:saao}}\orcidlink{0000-0002-6518-781X}
}

\institute{Cosmic Dawn Center (DAWN)\label{addr:DAWN}
\and
Niels Bohr Institute, University of Copenhagen, Jagtvej 128, DK-2200, Copenhagen N, Denmark\label{addr:jagtvej}
\and Trottier Space Institute at McGill and Department of Physics, McGill University, 3600 rue University, Montreal, Québec, H3A 2T8, Canada\label{addr:mcgill}
\and
South African Astronomical Observatory, P.O. Box 9, Observatory 7935, Cape Town, South Africa\label{addr:saao}
\and
Finnish Centre for Astronomy with ESO, FINCA, University of Turku, Turku, FI-20014, Finland\label{addr:finca}
}

\date{Received date /
Accepted date }

\abstract{
    The spectral features in the optical/near-infrared counterparts of neutron star mergers (kilonovae, KNe), evolve dramatically on hour timescales. To examine the spectral evolution we compile a temporal series complete at all observed epochs from 0.5 to 9.4\,days of the best optical/near-infrared (NIR) spectra of the gravitational-wave detected kilonova AT2017gfo. 
    Using our analysis of this spectral series, we show that the emergence times of spectral features place strong constraints on line identifications and ejecta properties, while their subsequent evolution probes the structure of the ejecta. 
    We find that the most prominent spectral feature, the 1\,\micron P~Cygni line, appears suddenly, with the earliest detection at 1.17\,days.
    We find evidence in this earliest feature for the fastest kilonova ejecta component yet discovered, at 0.40--0.45$c$; while across the observed epochs and wavelengths, the velocities of the line-forming regions span nearly an order of magnitude, down to as low as 0.04--0.07$c$. 
    The time of emergence closely follows the predictions for \srii, due to the rapid recombination of \sriii under local thermal equilibrium (LTE) conditions. The time of transition between the doubly and singly ionised states provides the first direct measurement of the ionisation temperature, This temperature is highly consistent, at the level of a few percent, with the temperature of the emitted blackbody radiation field. 
    Further, we find the KN to be isotropic in temperature, i.e.\ the polar and equatorial ejecta differ by less than a few hundred Kelvin or \(\lesssim 5\)\%, in the first few days post-merger, based on measurements of the reverberation time-delay effect.
    This suggests that a model with very simple assumptions, with single-temperature LTE conditions, reproduces the early kilonova properties surprisingly well.
    \newline 
        }

\maketitle
\section{Introduction}
The detailed spectra of the gravitational-wave detected kilonova GW\,170818/AT2017gfo \citep{Abbott2017b,Coulter2017} revealed the first spectroscopic identification of freshly synthesised \rprocess material in a binary neutron star (BNS) merger, \srii \citep{Watson2019}. Several \rprocess line-identifications have since been proposed in addition to \srii: \yii \citep{Sneppen2023b}, \teiii \citep{Hotokezaka2023}, \Laiii, and \Ceiii \citep{Domoto2022}. 
The time of appearance of these features can provide strong tests for line identifications, because the rapidly cooling ejecta undergoes quick transitions between ion states, at least under local thermal equilibrium (LTE) conditions \citep{Sneppen2023_rapid}. Tracking the subsequent evolution of the different spectral features also provides strong constraints for ejecta models including information on geometry and spatial abundance structures, potentially allowing us to `dissect' the kilonova. Spectra obtained with adequate temporal cadence can constrain different parts of the evolution at various times and wavelength ranges \citep[e.g.][]{Shappee2017,Andreoni2017,Nicholl2017,Chornock2017,McCully2017,Pian2017,Smartt2017,Tanvir2017}. 

To examine both the emergence and evolution of features requires a detailed early-to-late-stage spectral series with the broadest possible wavelength coverage, which no individual dataset contains. In this paper, we therefore track the spectral evolution of AT2017gfo, compiling and analysing the early spectra prior to 1.4\,days \citep{Shappee2017,Andreoni2017,Buckley2018}, the subsequent daily UV-NIR coverage using the VLT/X-shooter spectrograph \citep{Pian2017,Smartt2017}, as well as the space-based coverage from the \emph{Hubble Space Telescope} (\emph{HST}), which provide the only reliable data in the NIR telluric-dominated regions of this event \citep{Tanvir2017}. The combination of these datasets helps elucidate the physics of the spectral features in a way that no individual dataset has been able to.  

We present a summary of the datasets and consider the complementary information they contain in Sects.~\ref{sec:Overview} and \ref{sec:one_day_importance}. We analyse this compilation of all early published spectra of AT2017gfo in a consistent way to provide the blackbody temperature's evolution (Sect.~\ref{sec:temperature}) and as accurate a description as possible of when line features first appear (Sect.~\ref{sec:spectral_features}). We analyse the emergence time of the 760\,nm P~Cygni feature and its identification with \yii in Sect.~\ref{sec:760}. Similarly, we analyse the emergence of the strongest feature (at \(\sim1\)\,\micron) and check its consistency with the \srii interpretation in Sect.~\ref{sec:1.0mum}, and with a possible proposed association with \hei in a companion paper (Sneppen et~al.\ in~preparation). The NIR spectral features are also analysed, where we show that particularly the 1.4\,\micron feature displays an apparent P~Cygni-like spectral shape. We discuss the possible different origin of the NIR features compared to the optical lines (Sect.~\ref{sec:1.4micron}-\ref{sec:2micron}) and how this relates to the complementary nature of their velocity structure. 





\begin{figure*}
    \centering
    \includegraphics[width=0.95\linewidth,viewport=25 75 875 975 ,clip=]{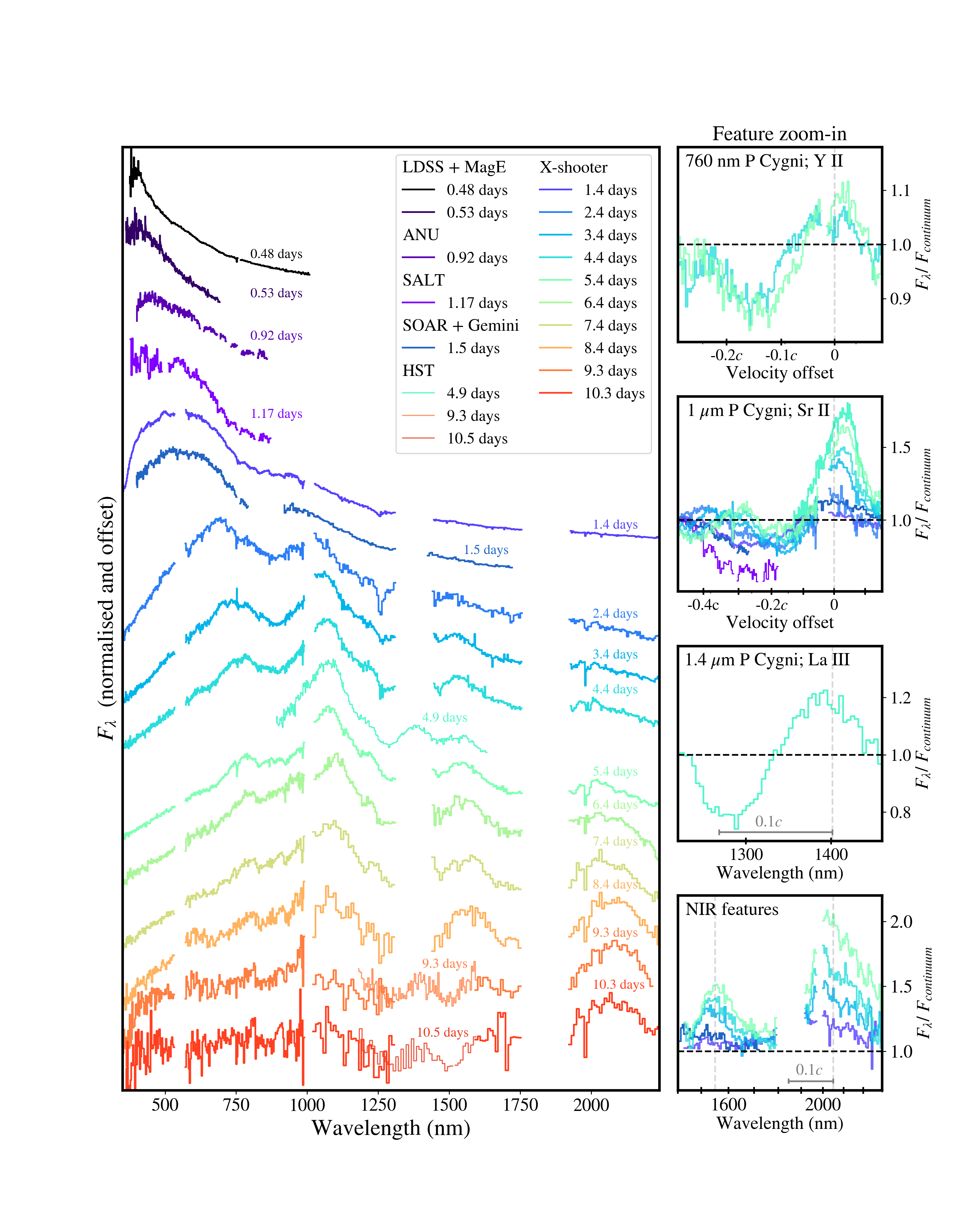} 
    \caption{Spectral series of the kilonova AT2017gfo showing the evolution over the first 10\,days post-merger. This spectral compilation of data includes the early spectra taken with the Magellan telescopes \citep{Shappee2017}, Australian National University 2.3-m telescope (ANU, \citet{Andreoni2017}), Southern African Large Telescope (SALT, \citet{Buckley2018}). From 1.4\,days post-merger, the X-shooter spectrograph at the European Southern Observatory's Very Large Telescope details the UV to NIR structure with a daily cadence \citep{Pian2017,Smartt2017}. To constrain the rapid evolution at early times, we also include the 1.5\,days Gemini FLAMINGOS-2 and SOAR spectra \citep{Nicholl2017,Chornock2017}. Towards intermediate and late-times \emph{HST} \citep{Tanvir2017} infrared spectroscopy additionally constrains the spectral structure in the telluric region around 1400\,nm. The spectral continuum is well-described as a blackbody in all photospheric epochs and even nebular-phase epochs given a blackbody temperature-evolution detailed in \ref{sec:temperature}. In the zoom-in sub-panels, individual spectral lines are highlighted at different epochs, including (1) the 760\,nm (2) 1.0\,\micron, and (3) 1.4\,\micron P~Cygni features, and (4) the NIR line-emission features. } 
    \label{fig:1}
\end{figure*}

\section{Overview of the data}\label{sec:Overview}
On 2017 August 17 at 12:41:04 UTC, the merging neutron star gravitational wave event GW170817 was detected and accompanied 1.74 seconds later by a short gamma-ray burst, GRB170817A \citep{Abbott2017b}. Following these triggers, a multitude of international observing campaigns searched for the optical counterpart, which was ultimately found in the host galaxy NGC\,4993 at 0.45\,days post-merger \citep{Coulter2017}. Over the subsequent hours, days, and weeks, a series of detailed spectra recorded the evolution of this rapidly evolving optical transient. In Fig.~\ref{fig:1}, we show the spectra selected for this analysis because of their 1) unique temporal coverage and then 2) their high signal-to-noise ratios and broad wavelength coverage. The times noted are the time post-merger in the restframe of the host galaxy NGC\,4993, $z_{\rm cosmic} \approx 0.01$ \citep{Hjorth2017,Howlett2020,Nicolaou2020}.

\textbf{\textit{0.48--0.53\,days post-merger:}} The first two spectra of AT2017gfo were obtained at 0.48 and 0.53\,days post-merger by the Magellan telescopes (with the Low Dispersion Survey Spectrograph, LDSS-3, and MagE spectrographs) at the Las Campanas Observatory, Chile \citep{Shappee2017}. These spectra display a featureless blue continuum well-modelled as a blackbody. The spectral coverage (LDSS-3: 425--1000\,nm, MagE: 360--700\,nm) does not include the blackbody peak at this early epoch, so the observed temperature (inferred from the spectral slope) is not as well constrained as in later epochs, but is likely around $10\,000$\,K \citep{Shappee2017,Waxman2018}. AT2017gfo faded measurably in the blue in the short  interval covered between these initial spectra, illustrating the rapid cooling of the transient. 

\textbf{\textit{0.92\,days post-merger:}} The third independent spectrum of AT2017gfo was taken at 0.92\,days with the Australian National University 2.3-m telescope at the Siding Spring Observatory \citep[ANU,][]{Andreoni2017}. The spectrum still appears largely featureless at this epoch. The optical wavelength coverage (320--980\,nm) now nearly covers the observed blackbody peak, indicating a significantly cooler transient, $T\simeq$~6000--7000\,K. We note the flux-calibration for the SED is somewhat uncertain, so the exact blackbody temperature inferred from the spectral slope is not very tightly constrained. However, the drastic decrease in temperature in the short time span since the earliest spectra is supported by photometric observations \citep[e.g.][]{Drout2017,Evans2017,Villar2017}, where interpolating the temperature at 0.92\,days would suggest $T = 6800\pm100\,K$.

\textbf{\textit{1.17\,days post-merger:}} The fourth spectrum in our series of AT2017gfo was taken at 1.17\,days with the Southern African Large Telescope \citep[SALT,][]{Buckley2018}. The optical wavelength coverage of SALT (380--900\,nm) reveals the first epoch where a pure blackbody continuum model poorly reproduces the data. Instead, sizeable absorption is seen redward of 650\,nm, which we argue in Sect.~\ref{sec:1.0mum} is the first appearance of the 1\,\micron P~Cygni feature. Due to the importance of the precise emergence time of the $\sim1$\,\micron feature, we have re-reduced the SALT spectrum using different standard stars in order to confirm that this observed feature is consistent and robustly constrained. We describe this re-reduction in Sect.~\ref{sec:SALT_red}. 

\textbf{\textit{1.4--10.5\,days post-merger:}} From 1.4\,days post-merger, AT2017gfo was monitored with nightly cadence with the X-shooter spectrograph mounted on the European Southern Observatory's Very Large Telescope \citep{Pian2017,Smartt2017}. 
The spectral range (330--2250\,nm) provides the first broad spectral coverage encompassing the ultraviolet (UV) through near-infrared (NIR). The earliest X-shooter epoch (1.4\,days post-merger) is remarkably well-modelled as a blackbody continuum in the UV, optical, and NIR \citep{Sneppen2023_bb} modulated by spectral features from strong \rprocess lines \citep{Watson2019,Gillanders2022}. Over the subsequent days and weeks the continuum fades, with the blackbody approximation increasingly poor, and becoming more dominated by the NIR emission, where a series of spectral components emerge more prominently. Many spectra were taken near-contemporaneously with the X-shooter datasets, e.g.\ NTT \citep{Smartt2017, Valenti2017}, Magellan \citep{Shappee2017,Nicholl2017}, VLT/MUSE \citep{Tanvir2017}, Gemini-South \citep{Chornock2017}, and SOAR \citep{Nicholl2017}. The various spectra are useful for validating the robustness of the spectral reductions, but the uniquely broad wavelength coverage and the high signal-to-noise ratio of the 8.2 meter VLT with the excellent throughput of X-shooter means that the X-shooter data overall provides the best constraints. Comparing across the approx.\ one hour difference in observing time between these observations can help in constraining the reverberation wave of recombining species as will detail in Sect.~\ref{sec:1micron_emergence}. We therefore use the SOAR and Gemini-South spectra \citep{Nicholl2017,Chornock2017} from 1.47\,days in addition to X-shooter as a compliment at this epoch.  
The data delivered by \emph{HST} \citep{Tanvir2017}, which provides a uniquely complementary spectral constraint in the telluric band around 1.4\,\micron (particularly using the Wide-Field Camera 3 Infrared channel with the grism G141). This region is difficult to confidently constrain from ground-based observations and therefore we include the 4.9, 9.4 and 10.6\,day post-merger \emph{HST} spectra in our data-series.

For consistency in comparing spectra, we apply a single Galactic dust extinction correction, using the reddening law of \citet{Fitzpatrick2004}, with $R_V = 3.1$ and $E(B-V)=0.11$. This $E(B-V)$ is the uncertainty weighted average of the \citet{Green2018} and \citet{Planck2016} constraints on the Galactic dust extinction along the line of sight. Similar to all previous reductions, the dust extinction at the location of AT2017gfo in the host galaxy is assumed to be negligible. For detailed arguments on the dust extinction see \citet{Sneppen2023A&A}. 

\begin{figure}
    \includegraphics[width=\linewidth]{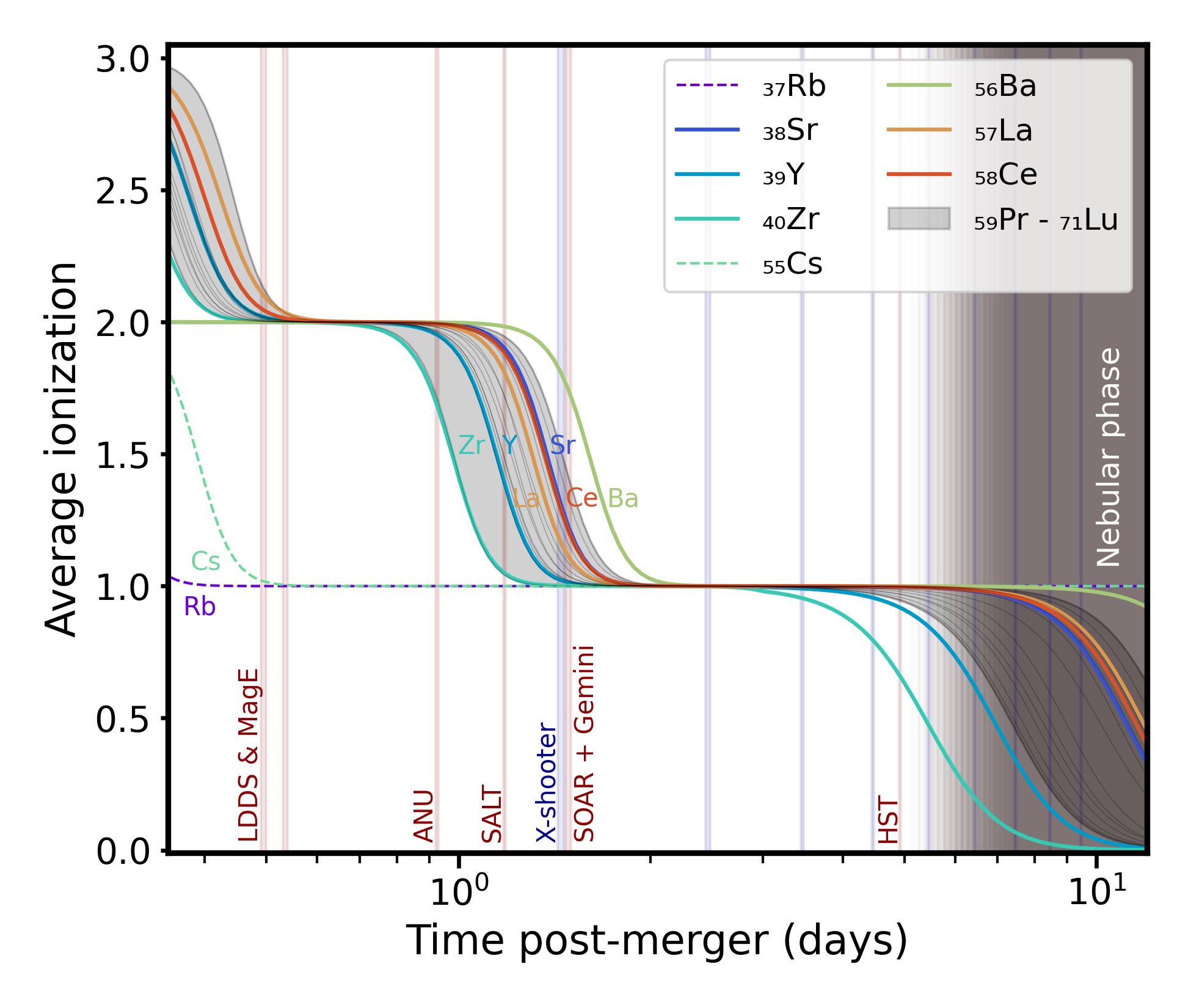} 
    \caption{Average ionisation inferred for the \rprocess elements in groups I--III. These elements are likely to produce the strongest lines under LTE conditions. We show them as a function of observed time post-merger. The ionisation states are calculated using the Saha equation, where we assume electron density, $n_e(t) = 10^8$\,cm$^{-3} ( t/1.5\,\textrm{days} ) ^{-3}$ and the blackbody temperature evolution found for AT2017gfo \citep{Sneppen2023}. The blue and red vertical regions indicate respectively the observations with X-shooter and other instruments (see the spectral series in Fig.~\ref{fig:1}). At certain temporal intervals, particularly at around 1\,day post-merger, the \rprocess elements dominating the opacity make rapid transitions between ionisation states. 
    For $t \gtrapprox 5$ days post-merger, NLTE-effects are likely to become important, implying that neutral, single, and double ionised species of these elements may be present \citep{Pognan2022}. } 
    \label{fig:ionization}
\end{figure}

\section{The one day transition and the importance of early spectra}\label{sec:one_day_importance}
To predict the expected emergence times of features, we show the average ionisation states for different \rprocess elements as a function of time under LTE conditions in Fig.~\ref{fig:ionization}. The elements include the \rprocess elements on the left side of the periodic table (i.e.\ groups I--III of periods 5 and 6, including the lanthanides). We include $_{37}$Rb and $_{55}$Cs for comparison, otherwise these are the elements which produce the strongest lines due to their few valence electrons, low-lying energy-states and small partition functions (i.e.\ Sr, Y, Zr, Ba) and the elements with high optical/NIR opacity with many low-lying states (i.e.\ the lanthanides). Both of these latter sets of elements display a rapid change in ionisation around one day post-merger.



For Fig.~\ref{fig:ionization}, we assume the temperature-evolution inferred from the Doppler-corrected blackbody continuum  \citep{Sneppen2023} with a homologously expanding (and thus diluting) electron density, $n_e(t) = 10^8$\,cm$^{-3} (t/1.5 \ {\rm days})^{-3}$ \citep{Sneppen2023_rapid}. At early times the temperature is cooling dramatically with $T \propto t^{-\alpha}$ with $\alpha \in [0.5;0.8]$ \citep[see Table \ref{table:table},][]{Drout2017,Waxman2018,Sneppen2023_rapid}, which is why the abundance of lowly ionised species rapidly transition from being a negligible population to the dominant species \citep[and thus entering the domain of new prominent spectral features, such as discussed in the models in][]{Tak2024}. Specifically, this sensitivity in LTE to temperature/time follows from the Saha equation, where the relative density of ions $n_{\rm i+1}/n_{\rm i}$ (with the subscript indicating the state of ionisation) is given by the electron-density, $n_e$, temperature, $T$, the ionisation energy, $\mathcal{X}_{\rm i}$, the partition functions for each ion, $Z_{\rm i+1}$ \& $Z_{\rm i}$, and the thermal electron de-Broglie wavelength, $\lambda_{\rm dB}$:
\begin{equation}
     \frac{n_{\rm i+1}}{n_{\rm i}} = \frac{2}{\lambda_{dB}^3} \frac{Z_{\rm i+1}}{Z_{\rm i}} \frac{1}{n_e} e^{-\frac{ \mathcal{X}_{\rm i} }{k_B T}} \label{eq:saha}
\end{equation}
The transition between ion states is especially sensitive to the temperature because of its exponential dependency. Due to the rapid drop in temperature over the first hours and days (see \ref{sec:temperature}), the ionisation transition is expected to be abrupt under LTE conditions. For \srii at 1.4\,days (where the $n_{\rm \sriii}/n_{\rm \srii}$ fraction is near-unity) the ionisation energy is $\mathcal{X}_{\rm 1} \approx 30 k_B T$, which implies a subtle decrease in temperature results in a large change in the \srii number density. As the ion recombines, the column density, and hence the optical depth, $\tau$, of the lines produced by the relevant species will change by orders of magnitude. This implies a drastic change in the transmission ($I = I_0 e^{-\tau}$) and thus a very rapid appearance or disappearance of features in LTE. 

Thus, the timing of the early spectral data from 0.5--1.5\,days post-merger (see Fig.~\ref{fig:ionization}) allows us to probe the ejecta before, during, and after this rapid recombination wave passes through the ejecta -- transitioning from doubly to singly-ionised states. The subsequent spectra constrain the evolving photospheric phase, the recession of the photospheric surface, and the transition to the nebular regime. The step-change between the dominant ionisation states under LTE followed by longer periods of relatively little evolution is a key observational difference to NLTE features. NLTE features are (by their nature) more smoothly and continuously evolving both during their (comparatively) slower formation-time and their subsequent evolution.

\begin{figure}
    \centering 
    \includegraphics[width=\linewidth,viewport=20 55 312 795 ,clip=]{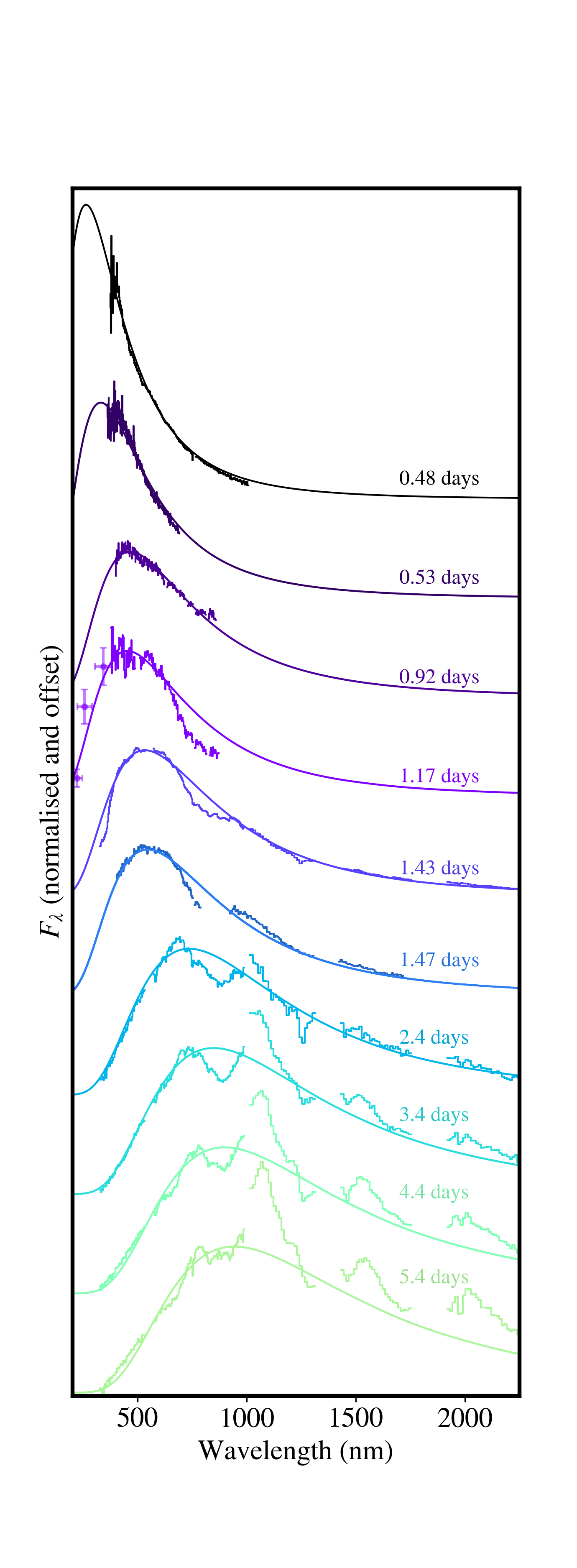}
    \caption{ Spectral series of the kilonova AT2017gfo with best-fit blackbodies overlaid. The temporal span here covers the photospheric epochs (i.e.\ spectra up to 5.4\,days post-merger), where a blackbody model approximates the continuum well. The blackbody temperatures and the spectral features are listed in Table~\ref{table:table} and  in Sect.~\ref{sec:spectral_features} respectively. } 
    \label{fig:spectral_temperature_series}
\end{figure}


\begin{table*}[t]
\caption{ Observed blackbody temperature required to match the spectral continuum at different epochs. At t $\lesssim1$\,day the blackbody peaks in the UV, while for every subsequent spectra both the Rayleigh-Jeans and Wien tails are constrained. In these subsequent spectra, dust extinction uncertainties contribute the dominant statistical uncertainty to the blackbody temperature. In Fig.~\ref{fig:spectral_temperature_series}, we show the best-fit blackbodies alongside the observed spectra.}\label{table:table}
\renewcommand{\arraystretch}{1.4} 
\centering
\begin{tabular}{@{}c @{\extracolsep{\fill}} ccc@{}}
\\ \hline \hline 
Time (days) & Telescope/Instrument  & \(T_{\rm obs}\) (K) & Publication 
\\ \hline 
0.48 & Magellan/LDSS        & \ \,\,\,\,\,\,\,\, 11\,000$^{+3400}_{-900}$ \,\,[1]  & \cite{Shappee2017} \\ 
0.53 & Magellan/MagE        & \,\,\,\,\,\,\,\,\, 9300 $\pm$ 300 \,\,[1]   & \cite{Shappee2017}  \\ 
0.92 & ANU/WiFeS         &  \,\,\,\,\,\,\,\,\, 6800 $\pm$ 200 \,\,[2]     & \cite{Andreoni2017} \\ 
1.17 & SALT/RSS        &  6400 $\pm$ 110   & \cite{Buckley2018}  \\ 
1.43 & VLT/X-shooter   &  5440 $\pm$ 60    & \cite{Pian2017}     \\ 
1.45 & VLT/X-shooter   &  5380 $\pm$ 60   & \cite{Sneppen2023_rapid} \\ 
1.47 & SOAR/GHTS, Gemini/FLAMINGOS-2 &  5330 $\pm$ 60          & \cite{Nicholl2017}, \cite{Chornock2017}  \\ 
2.42 & VLT/X-shooter   &  3940 $\pm$ 50    & \cite{Smartt2017}   \\ 
3.41 & VLT/X-shooter   &  3420 $\pm$ 40    & \cite{Pian2017}     \\ 
4.40 & VLT/X-shooter   &  3330 $\pm$ 40    & \cite{Smartt2017}   \\ 
5.40 & VLT/X-shooter   &  3070 $\pm$ 40   & \cite{Pian2017}     \\ 

\hline \hline
\end{tabular} \\ \, 


\begin{minipage}{\linewidth}
\footnotesize \textit{[1] These values of \(T_{\rm obs}\) are taken from \citet{Shappee2017} and remain unchanged as no spectral features have been noted in these epochs.}

\textit{[2] As the ANU spectrum does not strongly constrain the continuum shape, we report \(T_{\rm obs}\) for this epoch as constrained by preceding and subsequent photometry -- which suggests cooling from 7600 to 6600\,K between 0.7 and 1.0\,days post-merger \citep[see][]{Drout2017}. }
\end{minipage}
\end{table*}

\section{Blackbody Temperature Evolution}\label{sec:temperature}
The photometric \citep[e.g.\ Fig.~3 in][]{Drout2017} and spectroscopic \citep[e.g.][]{Pian2017} observations of AT2017gfo are remarkable well-described by a blackbody framework across the observed epochs. For instance the first X-shooter spectrum from 1.43\,days displays a temperature consistent at the 2\% percent level when inferred from the UV, optical or NIR separately (i.e.\ from wavelengths blueward and redward of the spectral peak) -- leaving little allowed territory for modifications to the blackbody \citep{Sneppen2023_bb}. Epoch-by-epoch estimates of the best-fit blackbody temperature required to match photometry and spectra respectively have previously been provided in \citet{Drout2017} and \citet{Waxman2018}. However, as these analyses predate the discovery and identification of the various spectral features, their modelling constituted a simple blackbody-fit to the whole spectral range. Not accounting for the modulations from spectral feature may bias their inferred blackbody temperatures. This bias can push the best-fit blackbody to colder temperatures (when neglecting modelling of absorption in the UV/optical in early spectra or not modelling NIR emission features) or hotter temperatures (when neglecting the prevalence of emission structures in the nebular phase from 1 week post-merger). 

Therefore, we provide an updated summary of the best-fit epoch-by-epoch temperatures in Table~\ref{table:table} and visualised in Fig.~\ref{fig:spectral_temperature_series}, where the spectral modelling now includes additional parameters for the features discussed in Sect.~\ref{sec:spectral_features}. Below we summarise the main uncertainties in estimating the blackbody temperature in the various observed periods.


In the first day post-merger, the flux peaks blueward of the optical spectroscopic coverage ($\lambda\lesssim400$\,nm). This means the temperature is only directly constrained from the spectral slope on the blue side of the blackbody in the earliest spectra. Fortunately, photometry from \textit{Swift} satellite (probing 200--400\,nm) at 0.6 and 1.1\,days constrains the UV flux, showing that it peaks at 300--400\,nm. Thus, while, for instance, the ANU spectrum at 0.92\,days does not independently provide strong constraints on the blackbody temperature due to its limited wavelength coverage, the preceding, contemporaneous and subsequent photometry puts strong bounds on the blackbody radiation field. 

When the spectral peak is constrained and during the photospheric epochs (\(1\)\,days \(\lesssim t\lesssim5\)--6\,days) the main uncertainty in the observed blackbody temperature lies in the exact Galactic dust extinction in the direction of the host galaxy \citep[see details of dust-extinction estimates for AT2017gfo in][]{Sneppen2023b}. Based on this work, we use $E(B-V)=0.11$ which is the weighted average of \cite{Green2018} and \cite{Planck2016}, and conservatively use a 15\% uncertainty including the absolute uncertainty in this direction. In addition, we also adopt a 10\% uncertainty on the $R_V$, based on the standard deviation of $R_V$ values found along different Milky Way sight-lines, which we propagate throughout the uncertainty estimates of this analysis. This total dust-extinction uncertainties implies as much as 5\% inherent uncertainty in the relative flux at 500\,nm relative to 1\micron, which corresponds to a 1--2\% uncertainty in the temperature inferred solely from the spectral shape and location of the spectral peak. We note, as the dust-correction will be the same for all epochs, the blackbody temperature uncertainty is correlated between epochs.

Moving into the nebular epochs (from 5--6\,days post-merger), the systematic uncertainties associated with identifying what constitutes the continuum and what constitutes nebular emission features will likely come to dominate. By 10.4\,days post-merger comparable flux is contributed by the emission features as from the underlying continuum -- implying any spectral continuum modelling is fraught with uncertainty. Further complicating constraints, the inferred spectral peak has shifted redward of 2.25\,\micron by this time and thus beyond the spectral regime constrained by X-shooter. It is noteworthy that NIR follow-up from \emph{JWST} at 1--5\,\micron of the recent GRB230307A (interpreted as being due to a BNS merger) is well-modelled as a blackbody continuum at $\sim$670\,K at 29\,days post-merger \citep{Levan2024}. This would tentatively suggest the blackbody continuum in kilonovae may even persist in some form towards late times. 

\section{Observed spectral features}\label{sec:spectral_features}
In the following, we consider each of the major spectral features, their emergence-time, velocity, and evolution. The elements associated with the proposed line identifications span the first and second \rprocess peaks and could potentially originate from different mass ejection channels and be distributed spatially inhomogeneously, both azimuthally and radially. This would be encoded in the emergence time, the line profiles, and in different velocities for each individual observed feature. In \citet{Sneppen2023}, we showed that the line-of-sight velocity from the \srii P~Cygni profile is highly consistent with the velocity inferred from the blackbody normalisation. This indicates that the outer ejecta (which defines the early photosphere) is highly symmetric and, potentially, that the light \rprocess element distribution may be as well. However, it is still unclear how constraining this velocity-consistency is for the underlying ejecta structures, with the simulation presented in \citet{Collins2024} showing that highly symmetric ejecta is not necessarily required to produce near-spherical line-forming regions. In \citet{Sneppen2023b}, we extended this analysis to the newly discovered 760\,nm yttrium P~Cygni profile, showing that it traces the velocity-structure and evolution of its strontium counterpart. In addition to these light \rprocess elements, the series of NIR features noted in previous work \citep{Pian2017,Smartt2017,Watson2019} has been linked to the second \rprocess peak elements \citep{Domoto2022}, but their velocity structure and their evolution has not yet been analysed within this framework. In this paper, we therefore use our spectral compilation to extend the velocity constraints to a broader temporal span including the NIR features. 

\begin{figure*}
    \includegraphics[width=\linewidth,viewport=-10 20 625 535 ,clip=]{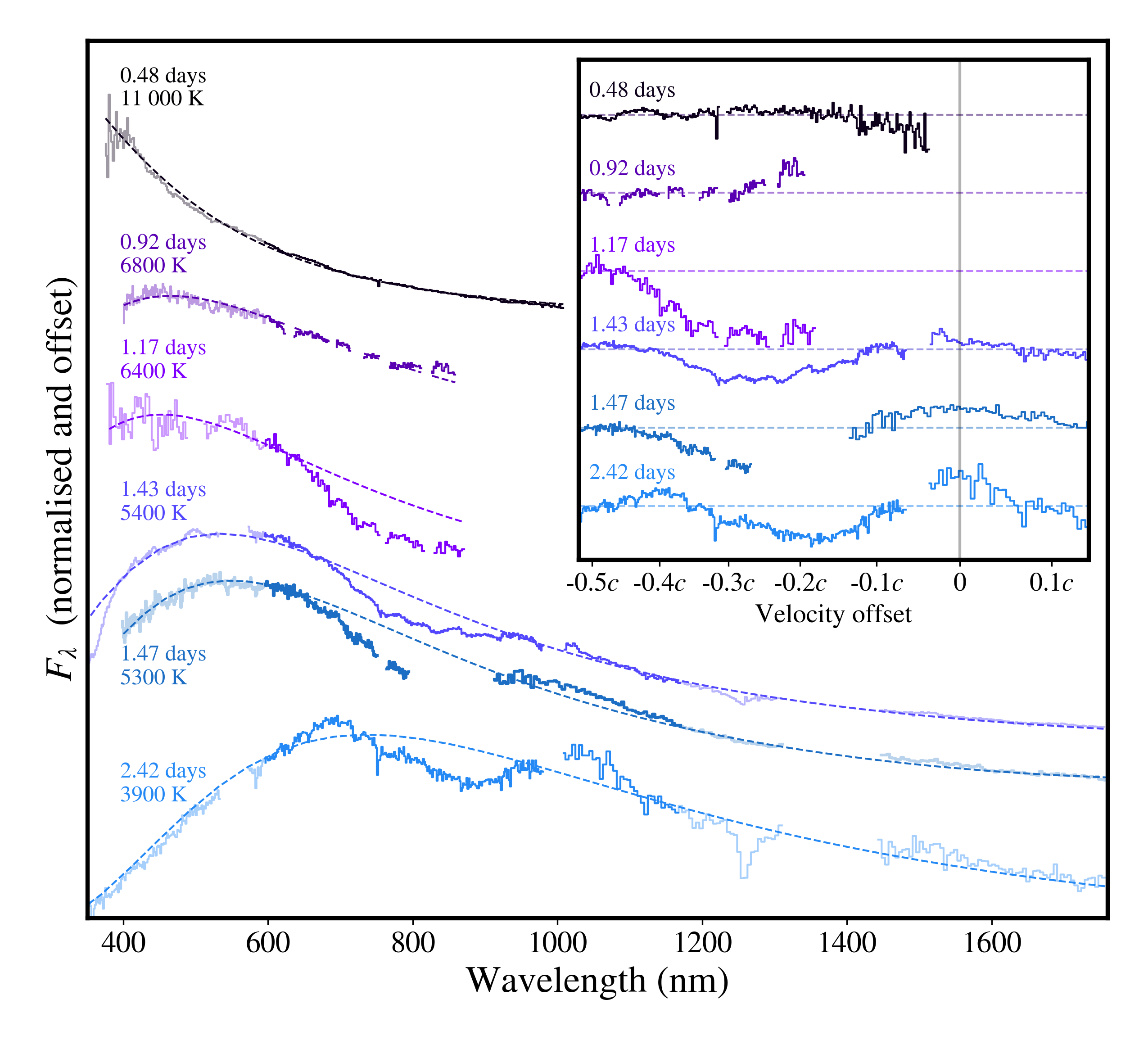} 
    \caption{Emergence of the 1\,$\mu$m P~Cygni feature in time. The selected spectra are from 0.48\,days \citep[LDSS,][]{Shappee2017}, 0.92\,days \citep[ANU,][]{Andreoni2017}, 1.17\,days \citep[SALT,][]{Buckley2018}, 1.43\,days \citep[X-shooter,][]{Pian2017}, 1.47\,days \citep[SOAR, \cite{Nicholl2017} and Gemini-South,][]{Chornock2017} and 2.42\,days \citep[X-shooter,][]{Smartt2017}. The spectra before 1\,day contain no strong deviation from the blackbody continuum. However, between 0.92 and 1.17\,days strong absorption appears, while the bulk emission begins to emerge with a delay of around 10\,hours. We note that this is the predicted emergence time-frame for both emission and absorption (and the early spectral shape) for a feature produced by LTE \srii due to a recombination wave passing through the ejecta \citep{Sneppen2023_rapid} -- and is inconsistent with a \hei\,$\lambda 1083.0$\,nm interpretation (see Sneppen et~al.\ in preparation). That the temperature-evolution constrained from the blackbody spectra predicts the formation time, indicates that the ejecta is close to LTE conditions and that the radiation blackbody temperature is a good probe of the ejecta ionisation temperature.} 
    \label{fig:sr_emergence}
\end{figure*}

\subsection{P~Cygni feature at 760\,nm  -- \yii?}\label{sec:760}
 The 760\,nm P~Cygni feature emerges 3--4\,days post-merger and, in terms of central wavelength (constrained by the emission peak) and the prominence of the feature (relative to its stronger \srii counterpart), is well reproduced by the 4d\(^2\)--4d5p transitions of \yii \citep{Sneppen2023b}. At earlier times the larger velocities mean that the various lines of \yii blend together thereby concealing the characteristic P~Cygni spectral shape. Modelling with the TARDIS code in \citet{Sneppen2023b} shows that a prediction of the yttrium identification is that the P~Cygni structure emerges clearly when the velocity drops to about 0.15--0.2\(c\), which is 3--4\,days after the merger. \citet{Tak2024} have now also identified the 4d\(^2\)--4d5p \yii transitions as likely to strongly affect line signatures around 760\,nm within their models. Other transitions to the 4d5p energy-level of \yii have been linked to the strong absorption observed at $\lambda \leq400$\,nm at early times \citep{Gillanders2022,Vieira2023}. Conversely, \citet{Pognan2023} find that the \yii transitions within their models are too weak to account for the observed 760\,nm features, and instead propose that in NLTE conditions \rbi may be able to produce a feature at roughly the right wavelength. However, as they use an uncalibrated theoretical line-list for yttrium which has incorrect wavelengths, placing these transitions at 1.1\,\micron (below the stronger \srii lines), it is unclear how constraining their conclusions on yttrium would be in a self-consistent model with the features placed at the correct wavelengths. As regards \rbi, no explicit fit of the observed feature is presented and their NLTE models cannot extend to the early times (2.4, 3.4 \& 4.4\,days post-merger) where the observed feature is first revealed and rises to prominence. Finally, the \rbi transition rest-wavelength (around 790\,nm) is 0.04--0.05c redward from the observed emission peak at early/intermediate epochs (in contrast to all other proposed line IDs -- \srii, \Ceiii, \Laiii, (and indeed \yii) -- which are consistent with the observed emission peaks at 1.0, 1.4 and 1.6 (and 0.76) \micron). 
This means the \rbi transition wavelength would be at odds with the spectral location and absorption/emission P~Cygni shape (assuming \srii, \Ceiii or \Laiii are reasonable proxies for the other observed features). In the later epochs the \rbi transition wavelength becomes more consistent with the observed feature, which 6--8\,days post-merger has evolved into what appears as a pure emission-like component.  

The velocity stratification of the feature 4.4--5.4\,days post-merger (see Fig.~\ref{fig:1}, first sub-panel) suggests a photospheric velocity around $0.15c$. This is in agreement with the velocity inferred from the 1.0\,\micron P~Cygni feature at this time. At later times, the feature evolves analogously to the 1.0\,\micron feature, reverberating into pure emission and fading away towards the more nebular epochs. Reverberation here means in the sense that the photons from different wavelengths of the P~Cygni profile are scattered at different physical times due to light travel-time effects. Thus as the line fades away it does so first in absorption and subsequently in emission. We will discuss this later-time evolution further in the context of the more prominent and, by extension, better constrained 1.0\,\micron feature.

\subsection{P~Cygni feature at 1.0\,$\mu m$  -- \srii }\label{sec:1.0mum}
The \srii 4p\(^6\)4d---4p\(^6\)5p triplet of strong lines producing the 1.0\,\micron P~Cygni feature were first proposed \citep{Watson2019} and later reproduced and extended with systematic analyses using different radiative transfer codes \citep{Domoto2021,Gillanders2022,Vieira2023} and with full 3D line-by-line opacity modelling \citep{Shingles2023}. The strontium mass required to produce the feature is compatible with what is produced in the dynamical and spiral-wave wind ejecta from merger simulations \citep{Perego2022}. Independent evidence for the \srii 4p\(^6\)4d---4p\(^6\)5p identification is that a large UV-absorption is observed at $\lambda \leq400$\,nm around 1.4\,days, which could be explained by \srii transitions from the ground-state to 4p\(^6\)5p \citep{Watson2019}, or similar lines from nearby and co-produced elements: \zrii \citep{Gillanders2022} and \yii \citep{Vieira2023}. Further corroboration of the \srii interpretation is the existence of a 760\,nm P~Cygni feature due to yttrium, discussed in the previous section \citep{Sneppen2023b}. 

An alternative interpretation has been proposed, that the 1.0\,\micron P~Cygni feature could originate from the neutral He\,{\sc i}\,$\lambda$1083.3\,nm line (1s2s-1s2p transition) which, in NLTE conditions could produce a feature at a similar wavelength \citep[see][]{Perego2022,Tarumi2023}, while \srii would be depopulated to higher states of ionisation. In that case, the apparently corroborating features of \srii, \yii, and \zrii at other wavelengths would need to be explained through different line-identifications with other transitions. However, as we show in Sneppen et~al. (in preparation), a \hei interpretation is not compatible with the emergence-time and subsequent spectral evolution of the observed 1\,\micron feature. This is contrary to the situation for \srii, as we discuss below.

\subsubsection{Emergence of the 1.0\,\micron P~Cygni feature}\label{sec:1micron_emergence} 
A key way to robustly discriminate between these, and indeed other proposed identifications, is to analyse the temporal properties and evolution of the feature. That is, in LTE, a \srii feature is predicted to begin its emergence only at a specific time, as the temperature is falling rapidly and thus the dominant ionisation level will also quickly change (see Fig.~\ref{fig:ionization}). As mentioned above, this is in contrast to  a helium interpretation, for example, which conflicts with the first appearance and the subsequent evolution of the feature. In the following, we show that the \srii feature's early evolution \citep[as per the predictions in][]{Sneppen2023_rapid} agrees with the observations. 
Previous analyses have focused on the X-shooter spectra, but the earlier spectra taken with the Magellan telescopes \citep{Shappee2017}, ANU \citep{Andreoni2017} and SALT \citep{Buckley2018} allow us to constrain the observed feature's earliest appearance. 



Given the blackbody temperature recovered in the X-shooter epoch~1 spectra (\(\sim1.4\)\,days post-merger), the expected ejecta electron densities, and the observed cooling-rate, we argued that a \srii P~Cygni feature should first start forming in absorption around $0.8$--$1.0$\,days post-merger -- with a significant absorption present by around $1.2$--$1.3$\,days post-merger \citep{Sneppen2023_rapid}. 

This timeframe is observationally probed by the early spectra (see Fig.~\ref{fig:sr_emergence}). The early spectra at 0.49, 0.53 and 0.92\,days all display a near featureless blackbody-like continuum. In contrast, the 1.17\,day SALT spectrum continuum is depressed by absorption in the red part of the optical ($\lambda \gtrapprox 650\,$nm for a blackbody with a temperature $T_{\rm obs} \approx 6400$\,K inferred from the location of the spectral peak). This is the first appearance of the 1.0\,\micron P~Cygni feature, consistent with the expectation from recombination of \sriii to \srii (Fig.~\ref{fig:ionization}). The observed absorption is a robust feature located in the middle of the wavelength coverage of SALT's prime focus Robert Stobie
spectrograph. The feature is also highly statistically significant: a blackbody continuum model, which adequately describes the spectrum below 640\,nm, describes the spectral region at 640--680\,nm poorly with a \(\chi^2\) null hypothesis probability of \(\sim10^{-26}\) when this region is added to the fit, and can thus be rejected at 10$\sigma$. Interestingly, that the absorption reaches blueward to around $\lambda \approx 640\,$nm, which indicates that the line-forming region must extend out to quite high velocities: 0.40--0.45$c$ (given absorption from ejecta expanding along the line-of-sight).
This is the fastest kilonova ejecta component observed so far -- $0.05c$ faster than the fastest component ejecta found from the 1.0\,\micron P~Cygni feature at 1.43\,days post-merger \citep{Sneppen2023,Vieira2023arxiv}. 
While rapidly receding, this ejecta velocity is still within the velocity range predicted from numerical simulations of kilonova dynamical ejecta \citep[e.g.][]{Bauswein2013}.

Due to the large time-delays across the ejecta, we expect to first observe the blueshifted outflowing material undergo an ionisation change, after which the more distant matter, producing the emission component of the P~Cygni feature, will be observed to undergo change in ionisation state with a delay of about 10\,hours. Thus, the earliest feature should form in very blueshifted absorption first, gradually evolving into a more pronounced absorption/emission P~Cygni feature. In reality, the first spectra with coverage of the \srii emission peak (X-shooter epoch~1) at 1.44\,days, does display only very weak emission, as expected due to this reverberation effect. At 1.47\,days (just 1\,hour after the epoch 1, X-shooter spectrum), the Gemini spectrum shows weak, but slightly stronger emission beginning to form \citep{Sneppen2023_rapid}. And consistent with this idea, by epoch~2 (2.4\,days post-merger), the feature has evolved into a clear P~Cygni shape with a pronounced emission component (Fig.~\ref{fig:sr_emergence}). 

That the observed temperature evolution of the blackbody spectra successfully predicts the line formation time, indicates that the ejecta are very near to LTE conditions and that the blackbody temperature is a good probe of the ejecta's ionisation/excitation temperature (we will revisit this further in Sect.~\ref{sec:LTE_validation}). 

\subsubsection{Evolution of the 1.0\,\micron P~Cygni feature}
The 1.0\,\micron P~Cygni feature, as the first clearly detected spectral feature, allows us to trace the evolution of the ejecta from 1.17\,days post-merger. The feature evolves in several key respects (Fig.~\ref{fig:1}, second sub-panel). 

First, the outer and inner (photospheric) velocities of the P~Cygni feature decrease with time \citep[see Fig.~\ref{fig:1} or][]{Watson2019}, because the line-forming region recedes deeper into the ejecta as the outer layers become optically thin. The photospheric velocity decreases from $0.28c$ (at 1.4\,days) to around $0.15c$ (at 5.4\,days). The optical depth of the absorption remains borderline optically thick throughout the early photospheric epochs with $\tau \in [1;3]$ from 1.17-3.4\,days, where-after it becomes optically thin and ultimately has faded away by 8\,days post-merger. 


Another important change of the feature is the increasingly prominent nature of the emission. This is caused by the line's time delay effect noted above and the rapidly cooling photosphere, but also the rapidly fading continuum; these effects creates spectral features that are quite sensitive to reverberation effects \citep{Sneppen2023_rapid}. 
This means that the amplitude of the P~Cygni absorption/emission is modulated not only by the changing ionisation state of the gas, but also by the changing strength of the underlying continuum, which itself rapidly fades with time. Modelling the changing continuum only produces a minor effect at times $t<2$\,days because of the slower flux evolution on the Rayleigh-Jeans side of the blackbody. However, this becomes pronounced at later times ($t>3$\,days), when the blackbody peak has shifted redward and into the feature, because the wavelength-dependent fading evolves rapidly near the blackbody peak \citep{Sneppen2023_rapid}. This effect is investigated in more detail in a future work (McNeill et~al., in prep). In the early nebular phase epochs (5.4--8.4\,days post-merger), the 1.0\,\micron feature fades away first in absorption and subsequently in emission. 



\begin{figure}
    \centering
    \includegraphics[width=\linewidth,viewport=18 17 380 280 ,clip=]{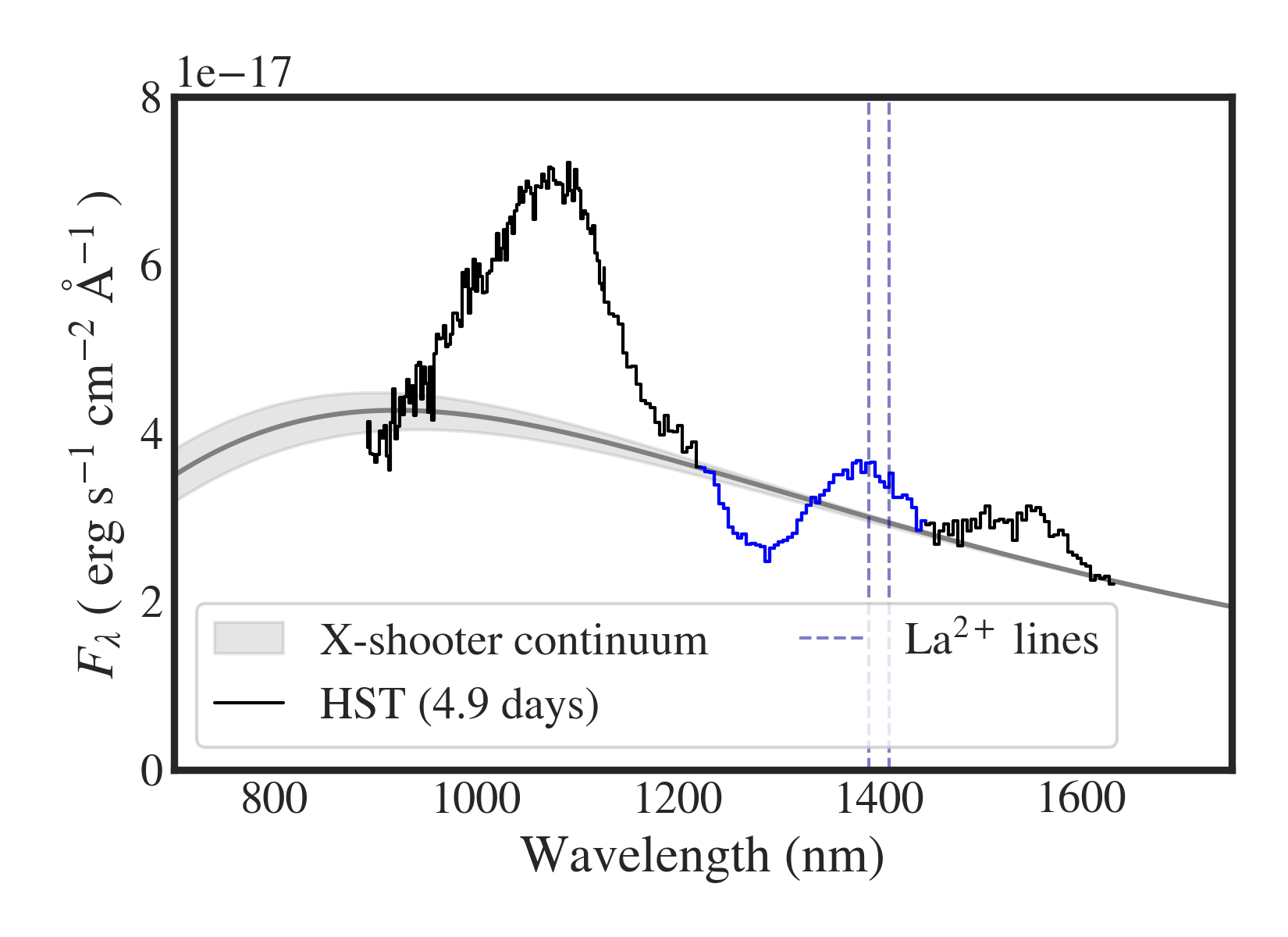}
    \caption{ HST spectrum of AT2017gfo at 4.9\,days post-merger from \citet{Tanvir2017}. The grey shaded region indicates the continuum constraints from the interpolation of the X-shooter spectra taken at 4.4 and 5.4\,days post-merger. Regardless of the exact continuum modelling, a P~Cygni absorption/emission feature around 1400\,nm is clearly visible. \citet{Domoto2022} suggested this feature could be due to \Laiii lines (here indicated with blue dashed lines). The characteristic velocity of this feature is relatively low, $v_{\rm ph} = 0.07 \pm 0.01 c$, but may be systematically biased by the strong nearby features (including the prominent reverberation emission peak around 1.1\,\micron). } 
    \label{fig:2}
\end{figure}

\subsection{P~Cygni feature at 1.4\,$\mu m$  -- \Laiii?}\label{sec:1.4micron}
A feature around 1.4\,\micron has been suggested as being due to the 5p\(^6\)5d---5p\(^6\)4f transitions in \Laiii \citep{Domoto2022}. The feature is poorly-constrained from the ground-based spectra as most of the profile resides within the telluric region. However, at 4.9\,days \emph{HST} spectra were taken of AT2017gfo \citep{Tanvir2017}, which revealed a P~Cygni-like feature with an emission peak around 1400\,nm and a blueshifted absorption around 1300\,nm (see third sub-panel in Fig.~\ref{fig:1} and Fig.~\ref{fig:2}). It is likely that this feature is also present in other epochs, potentially complicating the modelling of the 1.6\,\micron feature. Significant absorption from the continuum is noticeable around 1250\,nm as early as  2.4--3.4\,days post-merger (see Fig.~\ref{fig:1}, main panel). Notably, this is so late that the low blackbody radiation temperature would suggest \Laiii should be almost entirely absent under LTE conditions. This identification therefore seems to require either significant NLTE-effects at a few days post-merger, or that the line-forming region for this feature originates from an inner and potentially hotter ejecta which allows for doubly ionised lanthanides. We discuss this prospect further in Sect.~\ref{sec:vel-ion}. In the subsequent \emph{HST} spectra at 9.4 and 10.6\,days, a feature around 1.4\,\micron is still detected although on a continuum that is 5--10 times fainter \citep{Tanvir2017}. 

Given the relatively limited wavelength range of the \emph{HST} spectra, the continuum flux around this P~Cygni feature is difficult to ascertain directly. However, the spectral evolution is relatively minor between the temporally adjacent X-shooter spectra (at 4.4 and 5.4\,days post-merger), which allow strong modelling constraints on the continuum. Thus, within the range of  allowable blackbody models at 4.4 and 5.4\,days, we can model the continuum and, by extension, probe the nature of the 1.4\,\micron P~Cygni feature (see Fig.~\ref{fig:2}). Fitting the P~Cygni profile with this range of continuum models would suggest the photospheric velocity is \(v_{\rm ph} = 0.07 \pm 0.01 c\). This is a relatively low velocity -- around a factor of two lower than the contemporaneous light \rprocess P~Cygni features at 760\,nm (\yii, \citealt{Sneppen2023b}) and at 1\,\micron (\srii, \citealt{Watson2019}). This may suggest the optically thick region at these NIR wavelengths resides deeper within the ejecta (where the expansion velocity is smaller). 
However, the velocity inferred may also be biased by nearby features, e.g.\ the feature around 1.6\,\micron and the prominent reverberation emission peak around 1.1\,\micron, which could create a bias towards lower velocities. In addition, there appears to be a marginal feature at around 1.2\,\micron which is so far unaccounted for. 

\subsection{Feature at 1.6\,$\mu m$  -- \Ceiii?}\label{sec:1.6micron}
The 1.6\,\micron feature (see Fig.~\ref{fig:1}, fourth sub-panel) was interpreted by \citet{Domoto2022} as potentially being due to a combination of lines from \Ceiii, (predominantly 4f\(^2\)--4f5d transitions), which also produces several strong lines in the corresponding wavelength range of HR\,465 (the binary GY\,And), which is a metal enriched star \citep{Tanaka2023}. Conversely, Spectroscopic r-Process Abundance Retrieval for Kilonovae (SPARK) models in \citet{Vieira2023arxiv} show that \ceii can produce a spectral feature at a similar location in the NIR, though likely of lesser prominence. \cite{Gillanders2023} additionally notes the presence of nearby intrinsically weak [\ion{I}{ii}] lines which could contribute significantly to a feature towards late times. There is no statistically significant feature present at this wavelength in the 1.4\,day X-shooter spectrum. Curiously, the first weak emission has already begun to emerge around 1.47\,days in the Gemini spectrum -- seemingly in the narrow time-frame associated with the transition from doubly to singly ionised \rprocess line-forming species. Over the subsequent days the emission feature slowly becomes more prominent relative to the continuum. 
As in the case of \Laiii, the blackbody temperature and LTE conditions would suggest no significant \Ceiii population would be present in the photosphere from 2.4\,days onwards (see Fig.~\ref{fig:ionization}). Again, this would require significant NLTE effects at this time, or it may be possible that the NIR line-forming region originates from hotter ejecta. 

The feature and its mostly emission-like nature have been discussed previously \citep{Smartt2017,Watson2019,Sneppen2023,Gillanders2023}. The central wavelength of this feature shifts redward with time, starting at 1.52\,\micron (2.4\,days post-merger) and shifting to 1.58\,\micron by 7--10\,days post-merger \citep{Gillanders2023}. The velocity-width of this feature does not change substantially over time when modelled as a Gaussian emission feature, with a $\sigma$-width of only around 0.04--0.05\(c\). This is very narrow compared to the velocity measurements of other features, including even the redder 2.1\,\micron emission feature, which at face value appears to be broader, and which we tackle next. We note that it may be possible this is a full (and broader) P~Cygni absorption/emission feature modified by possible surrounding features. However there is no very clear signature of such a hypothesis, as the corresponding absorption would reside within the telluric regions and be highly degenerate with the 1.4\,\micron P~Cygni feature emission peak.  


\subsection{Feature at 2.1\,$\mu m$  -- {\rm [\teiiinospace]}?}\label{sec:2micron}
The 2.1\,\micron feature (see Fig.~\ref{fig:1}, fourth sub-panel) is well-modelled as pure emission and was interpreted as a [\teiiinospace] fine structure line by \citet{Hotokezaka2023} when modelling the spectra from 7.5--10.5\,days post-merger. \citet{Gillanders2023} also identified [\teiiinospace] as a plausible line-forming species, but further argued that due to the width of the line, the feature would likely have to be produced by two separate lines to match the narrowness of the 1.6\,\micron feature width. This feature (or a similar feature at more or less the same wavelengths) is also statistically significant in earlier epochs. It is relatively minor in the first spectra with NIR coverage (X-shooter epoch~1, 1.4\,days post-merger), grows in prominence with time (equalling the continuum at 5.4\,days post-merger) and in the later, probably more nebular phase, becomes the most prominent feature. The central wavelength drifts redward with time, starting at 2.0\,\micron at 1.4\,days but shifting to 2.1\,\micron by 10\,days post-merger. 

A feature at a similar wavelength has also been detected in late \emph{JWST} spectroscopy associated with a long-duration GRB and argued to be a kilonova from a merger \citep{Levan2024,Yang2024}. That feature was also tentatively associated with [\teiiinospace] \citep{Levan2024,Gillanders2023}, and may possibly have substructure, indicative of more than a single line.

Despite the considerable evolution in prominence of this feature in AT2017gfo, the velocity-width of the features remains largely constant at around $0.07c$ throughout all epochs from 1.4--10.4\,days post-merger. If the origin is several lines at different wavelengths this would naturally suggest an even lower expansion velocity, which as argued in \citet{Gillanders2023} could lower the velocity to be consistent with the measured width of the 1.6\,\micron feature. This velocity is particularly striking in two respects. 
First, in early epochs the velocity-width of the 2.1\,\micron feature is significantly smaller than the velocity inferred from the continuum, the 760\,nm P~Cygni feature, or the 1\,\micron P~Cygni feature, while it is consistent with the 1.4\,\micron feature (the reddest of the P~Cygni features) from the \emph{HST} spectra. This suggests the \rprocess elements creating this feature may not be co-located spatially with the light \rprocess elements producing the optical P~Cygni features. This either requires these lines to be located spatially near the equatorial plane (where the line-of-sight velocity is small due to projection-effect for a near-polar observer) or deeper within the ejecta (visible due to the lower opacity at longer wavelengths).
Second, the feature's velocity-width does not evolve significantly. If the elements producing the feature were confined to the merger's equatorial plane (which could explain the smaller projected velocity compared the optical P~Cygni features), then the relative velocity should still decrease as the line-forming region recedes inwards. If the line-forming region is located radially deeper in the ejecta, this non-evolution still requires a constant velocity-width of the line forming region. At the same time the observed drift of the central wavelength requires that the line-forming region is initially offset from the merger centre or perhaps that reverberation-effects are significant. Ultimately, modelling both a drifting central velocity and an invariant velocity-width seems difficult to reproduce and requires further study.  

\subsection{Minor and tentative features}\label{sec:tentative}
In the high S/N X-shooter spectrum from 1.4\,days, a few smaller, narrower spectral ``wiggles'' are noticeable in the absorption of the P~Cygni feature with the most noticeable peaks around 780\,nm and 860\,nm. These spectral features are persistent and appear in every single exposure taken throughout epoch~1 \citep{Sneppen2023_rapid}. Similar wiggles are seen in coincident MUSE \citep{Tanvir2017} and NTT \citep{Smartt2017} spectra, suggesting they may not be artefacts of the processing pipeline, but potentially the contribution of less prominent transitions yet to be identified. Such spectral ``wiggles'' with their smaller inferred (projected) velocity-scales could potentially follow from ejecta expanding equatorially given the polar inclination-angle. 

Alternatively, the positive and negative interference of various lines centred at different wavelength could in conjunction produce spectral wiggles with a somewhat smaller velocity/wavelength-width than that of any individual line. Indeed, these minor wiggles have  counterparts at similar wavelength in the subsequent spectra --  the 780\,nm wiggle has a clear spectral counterpart in the \yii 4d\(^2\)--4d5p transitions, which produces the 760\,nm P~Cygni feature from 3.4--6.4\,days when the characteristic velocities decrease and the lines have deblended. A minor feature around 860\,nm may also be noticeable in intermediate epochs (4.4--7.4\,days post-merger) as a small peak resides between the 0.76 and 1\,\micron features (see Fig.~\ref{fig:1}, main panel). This weak feature is particularly poorly constrained across the photospheric epochs (especially in epoch~2) as it is located within the absorption valley of the 1\,\micron P~Cygni feature. Coincidentally, 860\,nm is quite near the rest-wavelength of the strong NIR \caii triplet. Even a minor Ca abundance, Ca/Sr ratio < 0.002 in mass fraction, could produce a feature of this prominence, which would remain within the calcium upper bounds for AT2017gfo argued in \citet{Domoto2021}. However, the width of the feature does not fit well with the lines identified with Sr and Y, and if the feature really were to be due to \caii, the modelling is complex and outside the scope of this paper.

At 1.2\,\micron, a weak spectral wiggle is present in the spectra from 4--8\,days post-merger \citep[as previously noted by][]{Tanvir2017,Sneppen2023,Gillanders2023}. This feature is particularly poorly-constrained, as it resides between the prominent 1\,\micron reverberated emission peak and the absorption of the 1.4\,\micron P~Cygni feature. It is worth noting that in the metal enhanced star HR~465, which \citet{Tanaka2023} analysed for its lanthanide features, a few prominent (currently unidentified) lines are present around 1.2\,\micron, though they did not associate those lines with this feature.

\section{Implications for the ionisation-structure}
Having considered the various features and their observed structure, we now explore the information revealed when combining the various spectral identifications and the complete dataset. First, in Sect.~\ref{sec:LTE_validation}, we discuss how the 1\,\micron feature's emergence constrains the ejecta to be close to LTE conditions. Second, in Sect.~\ref{sec:vel-ion}, we discuss the complementary nature of comparing the optical and NIR velocity-structures. Lastly, in Sect.~\ref{sec:angular}, we discuss how the spatial distribution of elements may be constrained from the various P~Cygni features observed.

\begin{figure}
    \centering
    \includegraphics[width=\linewidth,viewport=20 20 405 370 ,clip=]{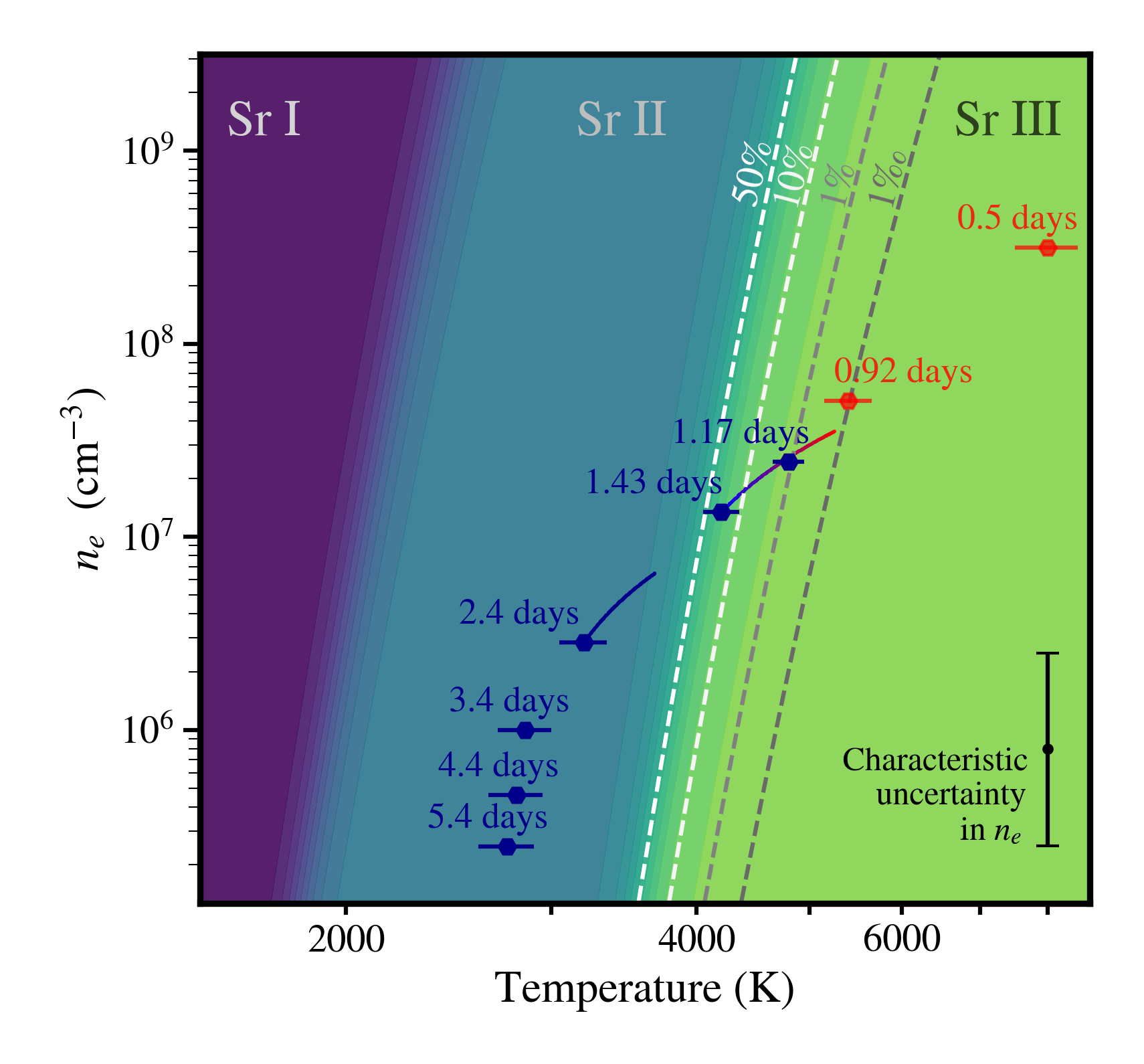}
    \caption{Ionisation degree of Sr in LTE as a function of temperature and electron density. We overlay the approximate locations in parameter-landscape of AT2017gfo with time, with lines indicating the evolution modelled as constraints from reverberation effects within a single epoch; the temperature errorbars are for the part of the ejecta with line-of-sight expansion \citep{Sneppen2023_bb} and the density assumes a spherically symmetric, homogeneous and homologously expanding high-velocity (\(0.2c \leq v \leq 0.3c\)) ejecta component of 0.01\,M\(_\sun\). The electron density can only be estimated to an order of magnitude, but regardless of the exact $n_e$ a rapid step-change in ionisation-state occurs over a narrow interval in temperature around 4500\,K. The 1\,\micron feature is not detected in observations prior to 1\,day (marked in red), but from 1.17\,days and forward the feature is present (marked in blue). } 
    \label{fig:Saha_eq_transition}
\end{figure}

\begin{figure}
    \centering
    \includegraphics[width=\linewidth,viewport=18 18 440 400 ,clip=]{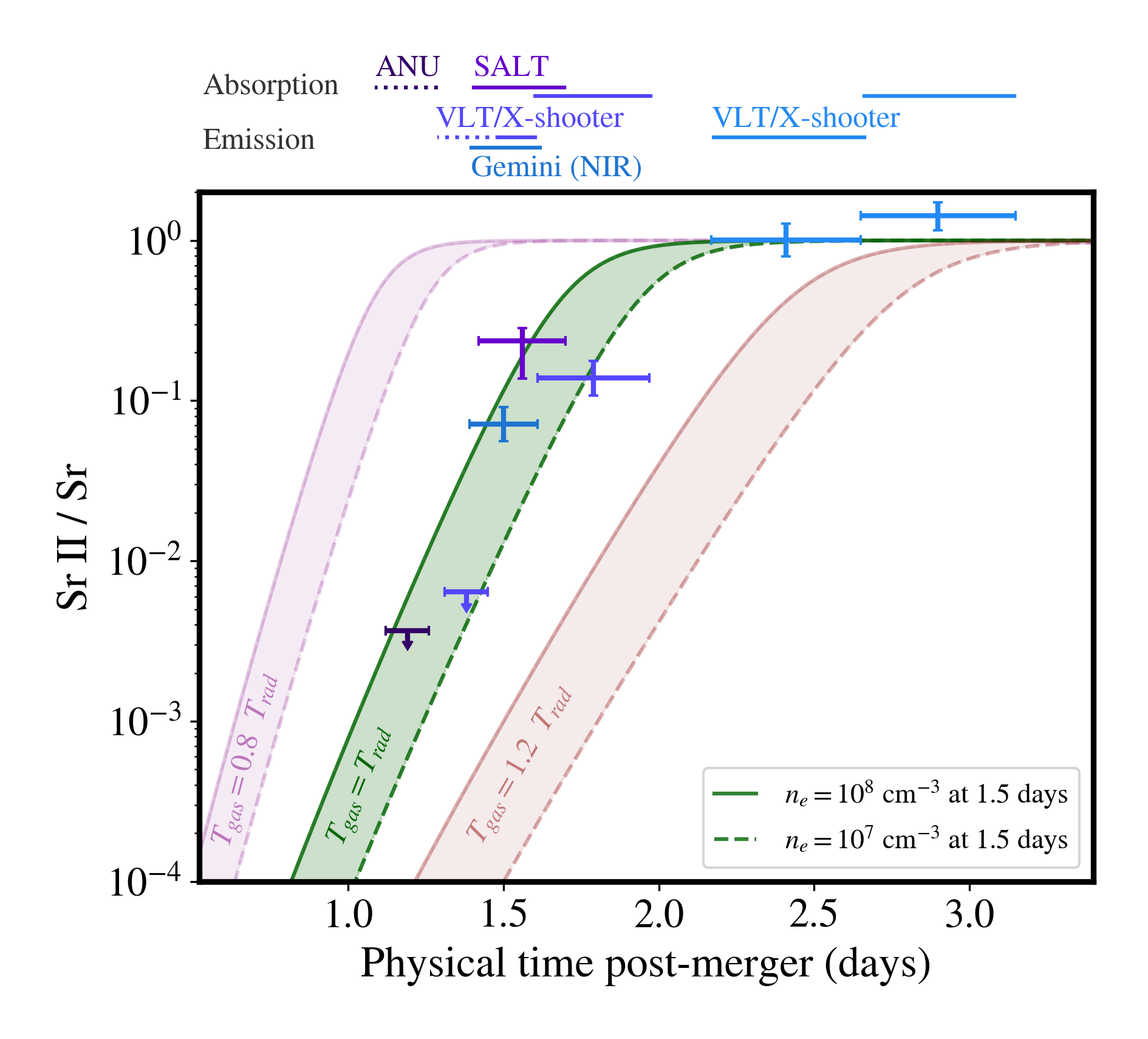}
    \caption{The fraction of \srii under LTE conditions as a function of time post-merger. On the upper x-axis, we illustrate the physical time of scattering probed by the 1\,\micron feature in absorption and emission by the various spectra from ANU, SALT, VLT/X-shooter and Gemini FLAMINGOS-2. Dotted lines indicate that no feature has clearly emerged (prior to $\sim$1.4\,days in both emission and absorption). 
    We indicates the required fraction of \srii needed to produce the optical depths observed in each spectrum (assuming \(M_{\rm Sr} = 6\times10^{-6}\)\,M\(_{\sun}\) homogeneously distributed across $0.2c<v<0.4c$, see main text). The ANU (X-shooter) and SALT (Gemini) spectrum indicate a rapid transition in absorption (emission) from an optically thin to optically thick regime around 1.3-1.5\,days. For the feature to emerge around 1.4\,days post-merger requires the ionisation temperature to be nearly identical to the radiation temperature inferred from the blackbody emission. For the feature to emerge at near-contemporaneous physical time in the polar and equatorial ejecta requires a near-isotropic temperature. } 
    \label{fig:Saha_formation}
\end{figure}

\begin{figure}
    \centering
    \includegraphics[width=\linewidth,viewport=5 5 475 460 ,clip=]{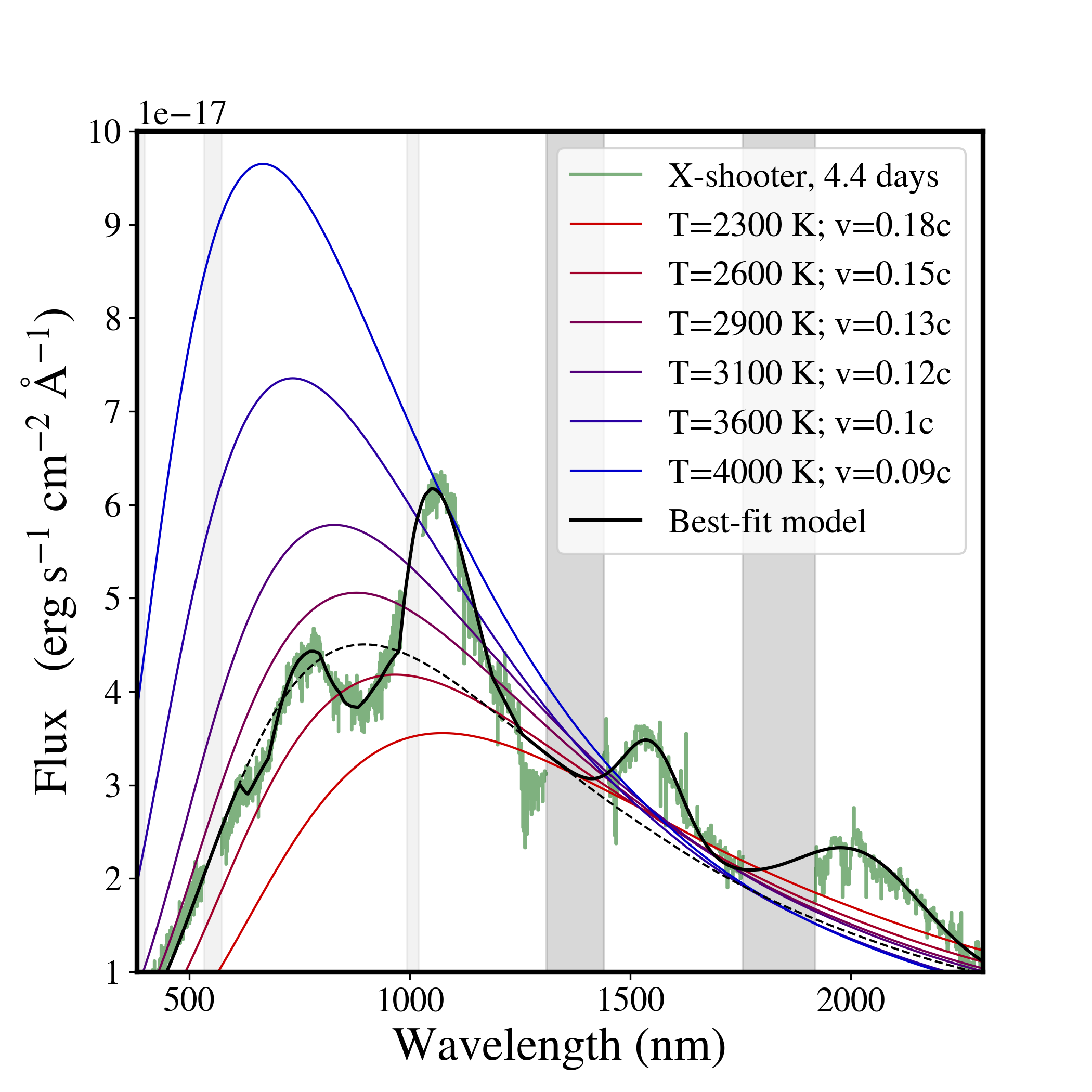} 
    \caption{Degeneracy between the blackbody temperature and the inferred emitting area (or velocity) for a given NIR continuum. The X-shooter epoch~4 spectrum is overlaid with blackbodies of various temperatures and areas which produce similar NIR flux levels. The observed NIR features can be modelled as P~Cygni or pure emission components, but make the exact underlying continuum difficult to estimate precisely. If the near-observed infrared features are produced by \Laiii and \Ceiii for LTE conditions, such species must reside in relatively hot ejecta. Hotter ejecta produces a higher NIR continuum flux, which would conversely require a smaller emitting area/velocity to match the observed continuum. While the optical P~Cygni features are consistent with the observed blackbody temperature and velocity, these NIR identifications would need higher temperatures and be spatially distinct from their optical counterparts.
    } 
    \label{fig:velocity_ionisation}
\end{figure}

\subsection{Quantifying the validity of LTE at early times}\label{sec:LTE_validation}
Perhaps the most striking aspect of the many early spectra taken of AT2017gfo is their relative simplicity. A simple blackbody empirically describes the continuum very well at early times \citep[e.g.\ the first X-shooter spectrum at 1.43\,days has percent-level consistency in the inferred temperature from the UV through the NIR, see][]{Sneppen2023_bb}. The spectral perturbations away from the blackbody are produced in the surrounding atmosphere by the LTE \rprocess species with the strongest individual transitions \citep[e.g.\ elements from the left side of the periodic table with low-lying energy-levels and small partition functions, see][]{Watson2019,Domoto2022}. While both the continuum and the spectral features are thus well-described by LTE populations at early times, it is still unclear how strong NLTE effects are permitted by observation.

By studying the emergence of features we can now begin to observationally constrain the allowed level of deviations from LTE conditions at early times. That is, the blackbody temperature not only predicts the dominant ionisation level (and thus which spectral features are present), but the temperature evolution accurately predicts the transition time between ionisation states (and thus when features form) as shown in Fig.~\ref{fig:Saha_eq_transition}. In Sect.~\ref{sec:1micron_emergence} we saw how, on an hour-by-hour basis, the spectral series allowed us to trace the formation of the 1\,\micron feature, first tracing the formation of the absorption component at around 1.0 day and subsequently (with a 10\,hour time-delay) the emission component, which started to form from around 1.4\,days. Both of these timeframes were predictions in \cite{Sneppen2023_rapid} due to an LTE recombination-wave from \sriii to \srii. This empirically validates that the radiation temperature of the blackbody is well-coupled to the ejecta temperature, because for every 100\,K of difference invoked between these temperatures the emergence time would have to change by around one hour. If we assume we don't know the electron density to an order-of-magnitude (i.e.\ propagate a total uncertainty in $n_e$ of 1.0\,dex), this corresponds to an uncertainty of 2\,hours in emergence-time towards either earlier or later formation. Within this level of uncertainty the emergence time in absorption and emission is as predicted for LTE, implying the difference between the excitation/ionisation and blackbody radiation temperatures must be within $\sim$200\,K at these times, i.e.\ consistent at the level of some percent. 

These strong constraints on the temperature are attainable because the timing of ionisation transitions is so strongly dependent on temperature and only relatively weakly affected by the electron density, as illustrated in Fig.~\ref{fig:Saha_formation}. Specifically, this shows the recombination-transition to \srii \ -- and, by extension, the formation-time of a \srii feature -- given Eq.~\ref{eq:saha}, typical KN electron-densities, and for various relationships between electron temperature and the blackbody radiation temperature. For comparison, we calculate the fraction of strontium required as \srii to produce the observed Sobolov optical depth $\tau=1$, where the strontium density is computed using the total strontium mass estimate \(M_{\rm Sr} = 6\times10^{-6}\)\,M\(_{\sun}\) \citep[as inferred from in the line-forming region at 2.4\,days,][]{Gillanders2023} homogeneously distributed between $0.2c<v<0.4c$. 
We have used the mass estimate at 2.4\,day as a benchmark as this is the first epoch where \srii is the dominant ionisation state, but note that there are large uncertainties in the exact strontium densities and indeed its radial structure. The y-axis errorbars in Fig.~\ref{fig:Saha_formation} thus solely represent the statistical uncertainty in the optical depth of the observed feature from each spectrum -- as modelled by the \srii 4p\(^6\)4d--4p\(^6\)5p lines, given LTE level-populations.

However, especially at early times the evolution of $\tau(t)$ is very constrained under LTE. Particularly, we must i) form a feature on a timescale of hours around 1.3 days, ii) maintain an optically thick line on a timescale of a few days, and iii) subsequently become optically thin and fade away at late times. This first aspect is observationally constrained in the polar ejecta by the ANU and SALT spectra, which are marked by no clear absorption ($\tau<0.08$; 3$\sigma$ upper limit) and a strong absorption ($\tau=2.2\pm 0.6$) respectively. In physical time, this represents an order of magnitude increase in the optical depth of the line from 1.3 to 1.5\,days. The optical depth of the emission feature can similarly be estimated using the P~Cygni framework under the assumption of pure scattering. In this case, the optical depth increases from $\tau<0.09$ (X-shooter; 3$\sigma$ upper limit) to $\tau = 0.8\pm 0.2$ (Gemini FLAMINGOS-2) from 1.43\,days to 1.47\,days. This highlights a similar rapid line formation in the equatorial ejecta, but now confined to the timescale of an hour due to the higher cadence around this period. 

It's interesting to note that X-shooter's NIR-arm data actually constrains the scattering of light (i.e.\ emission) at an earlier physical time than the SALT spectrum's optical light (i.e.\ the absorption), despite the former spectra being obtained 6\,hours \emph{after} the latter. 



\subsubsection{The temperature isotropy of the line-forming region}
We note that the fact that one can successfully predict both the formation time in absorption and (in the delayed) emission, accounting only for light travel time effects, indicates that there are not large temperature gradients from the polar to equatorial ejecta. Specifically it begins forming in absorption (i.e.\ the polar ejecta) between ANU and SALT spectra (indicating formation around 1.3--1.5\,days in physical time post-merger) and is beginning to form in emission around the time of X-shooter and Gemini (1.43--1.5\,days). This near consistency in emergence to within 2--3\,hours suggests the polar and equatorial line-forming region must reach the temperature for which \srii can first exist (i.e.\ \(T\sim 4500\)--5000\,K) at nearly the same physical time. This requires a consistency in ejecta temperature between the polar and equatorial ejecta within 200--300\,K, i.e.\ an angular consistency in temperature at the level of a few percent. 

This again illustrates that the remarkably simple framework of a single-temperature blackbody (as suggested by the spectral shape) may, surprisingly, capture the complexities of the ejecta. For comparison, in the state-of-the-art 3D models presented in \citet{Shingles2023}, large temperature anisotropies exist between the equatorial and polar ejecta -- at the level of up to 50\%. Such a class of model cannot produce the emergence of the 1\,\micron features observed in AT2017gfo. Thus, the discovery of these new observables measuring the ionisation temperature and the temperature anisotropy around ionisation transitions can inform and constrain future modelling of BNS mergers, specifically the thermalization of decay products and the radiative transfer in KNe. 

At intermediate times, complexities emerge as we may begin to see both single and doubly ionised species producing spectral features, which is the focus of our next discussion.

\subsection{ Velocity/ionisation complementarity }\label{sec:vel-ion}

\srii and \yii can only co-exist with the higher ionised species, \Laiii and \Ceiii, over a small range of temperatures at LTE due to the similar first and second ionisation energies (see Fig.~\ref{fig:ionization}). It is therefore unlikely that these species are co-spatial with their features appearing together across the wide temperature range probed at the different photospheric epochs. As argued above, if both optical and near-infrared identifications are correct, their co-existence in LTE would require their line-forming regions to be spatially separated by ejecta with different physical temperatures. The more highly ionised species would only exist, for example, in deeper (and thus likely hotter) parts of the ejecta. This outer and inner ejecta could plausibly be simultaneously probed by different wavelengths due to the expected decline in overall opacity towards the NIR. This spatial separation between species would map onto different expansion velocities, such that the velocity difference between the optical P~Cygni line profiles (from singly-ionised species) and the near-infrared features (doubly-ionised species) could be due to different ejecta temperatures. 

This interpretation can be tested for its self-consistency as the temperature must both be consistent with 1) the ionisation-level of the proposed species and 2) the continuum flux given the emitting area set by the photospheric velocity of the line and free expansion. 
In the UV-optical, the temperature and velocity are individually well-constrained by the spectral shape and location of the blackbody peak -- however in the Rayleigh-Jeans tail toward the NIR the temperature and emitting area can be varied along a line-of-degeneracy to get a similar continuum (Fig.~\ref{fig:velocity_ionisation}).
To produce a flux at \emph{one} specific wavelength there is a complete degeneracy between emitting area and temperature, but the NIR spectral \emph{slope} is sensitive to the temperature. For example, the NIR temperature is consistent with the UV/optical temperature at the percent level at 1.4\,days \citep{Sneppen2023_bb}. 
For doubly-ionised species to be present at typical electron densities (at 4\,days this is around $10^7$\,cm$^{-3}$) one would require the electron temperature to be around 4000\,K or hotter. This is much hotter than the optical-UV temperature at this epoch. However, such an ejecta temperature is broadly consistent with the NIR continuum flux for the observed velocity inferred from the 1.4\,\micron line. But, it is worth noting that these NIR-continuum temperature constraints for any fixed velocity are likely upper bounds, as they neglect modelling the contribution of fluorescence from all layers at greater radii, which requires detailed radiative transfer modelling, so the actual radiation temperature could be lower.

If these identifications of doubly-ionised species are correct, this diversity in the ionisation levels of species emerging at intermediate epochs would constitute the first observational hints of a multi-temperature ejecta/blackbody. That the more highly ionised species first appear (and increase in prominence) after the observed optical-UV blackbody temperature cools away from the species' dominant temperature-regime ($T \lesssim 4000$\,K) might at first be considered surprising, but could be explained by the relevant species initially being hidden within optically thick inner regions \citep[e.g.][]{Vieira2023arxiv}. 
On the other hand, the appearance of more highly ionised species in the NIR could be a distinctly NLTE phenomenon. Indeed the gradual emergence and continuously increasing prominence are key observational properties of NLTE features (in contrast to the LTE evolution). Towards late times ($t \gtrsim 5$\,days), NLTE effects may boost the abundance of higher ionisation species \citep{Pognan2022}. Finally, it may be that the identifications with more highly ionised species is simply not correct and we await the correct identifications of these lines. However, in that case, the apparently low velocities of these lines would still need to be explained.

\subsection{The angular distribution of line-forming species}\label{sec:angular}
The observation of several P~Cygni structures in the line-forming regions (see Fig.~\ref{fig:1}, right panels), requiring both absorbing and emitting components, provides hints of the spatial geometric distribution of the elements. As we are observing this object with a near-polar inclination angle \(\theta_{\rm inc} = 22\degree\pm3\degree\) \citep{Mooley2018,Mooley2022}, the blueshifted absorption and the rest-wavelength emission here closely correlate with the more polar and equatorial components of the ejecta respectively. 

With the emergence of the 1\,\micron feature as detailed in this compiled spectral series, we have further tightened the connection between \srii and the observed emission of the 1\,\micron P~Cygni feature. This requires a sizeable component of strontium (i.e.\ a first peak \rprocess element) to be located in the equatorial plane. Further, the emission-part of the 0.76\,\micron P~Cygni feature, again interpreted as due to first peak \rprocess species, \yii \citep{Sneppen2023b}, supports the interpretation of light \rprocess elements in the equatorial plane in addition to the polar (absorption) component. Such equatorially-distributed light \rprocess elements are certainly not a general expectation from BNS merger simulations, which typically produce light \rprocess ejecta principally in the polar direction \citep[e.g.][]{Kasen2015,Collins2023,Just2023}. Lastly, if the 1.4\,\micron feature is P~Cygni-like as argued in Fig.~\ref{fig:2} and the \Laiii interpretation suggested in \citet{Domoto2022} is correct, this would similarly suggest that second peak \rprocess elements are also spatially distributed in a manner which allows both absorption and emission. This would seemingly suggest a moderate abundance of lanthanides at and above the NIR photosphere in the polar ejecta -- in addition to the more generally expected equatorial (emission) component found often in simulations \citep[e.g.][]{Just2023}.

Thus, it is likely that the near-spherical distribution of \srii in AT2017gfo found in \citet{Sneppen2023}, which contrasts with predictions from simulations, is not exclusively an issue in strontium. 
However, the full landscape of these geometrical constraints have yet to be explored. The development of full 3D radiative transfer modelling coupled self-consistently to hydrodynamically simulated kilonovae, such as that pioneered in \citet{Shingles2023,Collins2024}, may provide an avenue to bridge the interpretability of BNS merger simulations and the observed spectra of AT2017gfo and future KNe. These models currently do not predict or reproduce P~Cygni structures due to the large spatial asymmetries of the various line-forming species from hydrodynamic simulations. 

\section{Conclusion}\label{sec:discussion}

We have compiled a detailed early-to-late spectral series of AT2017gfo to investigate the emergence and evolution of spectral features from freshly synthesised \rprocess elements. Combining these data has revealed:  
\begin{enumerate}
    \item The earliest chronological appearance of the 1\,\micron P~Cygni feature (in the SALT spectrum from 1.17\,days) reveals the fastest kilonova ejecta component yet discovered at $0.40--0.45c$. 
    
    \item The 1\,\micron emerges as blueshifted absorption first, gradually evolving into a more pronounced P~Cygni-like feature. This follows the predictions in \citet{Sneppen2023_rapid} on the exact emergence-time and early spectral shape from strontium undergoing a \textsc{iii}--\textsc{ii} recombination wave in LTE. The transition time between ionisation states empirically shows that the radiation temperature of the blackbody is well-coupled to the ionisation/excitation temperature of the particles themselves and observationally constraining the ejecta to be close to LTE.
    

    \item The observed spectral features at 0.76, 1.0, 1.4, 1.6, and 2.1\,\micron have been described and compared in their velocity structures. The two optical P~Cygni features are broadly similar in velocity over time, while their NIR counterparts appear to have lower velocities. This difference in velocities may relate to a difference in the proposed identifications for the optical (\srii, \yii) and NIR features (\Laiii, \Ceiii, \teiii), i.e.\ being respectively first vs.\ second \rprocess-peak elements and single vs.\ doubly ionised and hints at a possible spatial separation.

    \item The abundance of P~Cygni structures in the line-forming regions, requiring both absorbing and emitting components, indicates the spatial geometric distribution of the elements. Particularly, the \srii and \yii emissions requires a sizeable component of first peak elements to be located in the equatorial plane. Such an element distribution is in contrast with the typical spatial predictions from simulations. However, this is likely no longer exclusively an issue in \srii (or \yii), as these dual emission/absorption components are also seen for the 1.4\,\micron P~Cygni feature. Furthermore, that the recombination of \sriii to \srii in absorption and emission occurs at the same physical time (i.e.\ 1.3--1.5\,days post-merger) indicates a temperature in the line-forming-region around this time which is isotropic at the level of 5\% from the polar to the equatorial ejecta.

\end{enumerate}




\section*{Data and code availability}
As discussed, the spectral series presented in this paper is composed from a series of different observing programmes at various telescopes. We request that any use of the data whether compiled or re-reduced for this analysis includes appropriate citation to the original papers. 

Spectra from the 6.5\,m Magellan Telescopes (0.49 and 0.53\,days) located at Las Campanas Observatory, Chile, can be found at: \url{https://arxiv.org/abs/1710.05432}. The 0.92\,day spectra were obtained with the Australian National University (ANU) the 2.3\,m telescope located at Siding Spring Observatory. The 1.17\,day data obtained with the Southern African Large Telescope (SALT) under the Director’s Discretionary Time programme 2017--1-DDT-009, are available at \url{https://ssda.saao.ac.za} with the newly reduced spectra (e.g.\ with improved flux-calibration, seeSect.~\ref{sec:SALT_red}) now available at \url{https://github.com/Sneppen/Kilonova-analysis}. X-shooter data from European Space Observatory (ESO) telescopes at the Paranal Observatory under programmes 099.D-0382 (principal investigator [PI]: E.~Pian), 099.D-0622 (PI: P.~D’Avanzo), 099.D-0376 (PI: S.~J.~Smartt), which are available at \url{http://archive.eso.org} and WISeREP (\url{https://wiserep.weizmann.ac.il/}). The re-reduced X-shooter spectra examining evolution in sub-epoch exposures at 1.4\,days (derived in \cite{Sneppen2023_rapid}) are available from \url{https://github.com/Sneppen/Kilonova-analysis}. Gemini-south data were obtained at the Gemini Observatory (Program IDs GS-2017B-Q-8 and GS-2017B-DD-4; PI: Chornock) and are together with SOAR spectra available at \url{https://kilonova.org}. \emph{HST} observations were obtained using programs GO~14771 (PI: N.~Tanvir) and GO~14804 (PI: A.~Levan).

We use the implementation of the P~Cygni profile in the Elementary Supernova from \url{https://github.com/unoebauer/public-astro-tools} with generalisations to account for reverberation, time-delay effects and special relativistic corrections, see \url{https://github.com/Sneppen/Kilonova-analysis}. 

\section*{Acknowledgements}
The authors would like to thank Igor Andreoni, Steven Crawford, Jonathan Selsing, and Jesse Palmerio for sharing, introducing and clarifying the various datasets and data-reductions. We would further like to express our gratitude to Stuart Sim, Christine Collins, Luke Shingles, and Fiona McNeill for discussions on the spectral modelling. 

The Cosmic Dawn Center (DAWN) is funded by the Danish National Research Foundation under grant DNRF140. AS, DW, RD, and KEH are co-funded by the European Union (ERC, HEAVYMETAL, 101071865). Views and opinions expressed are, however, those of the authors only and do not necessarily reflect those of the European Union or the European Research Council. Neither the European Union nor the granting authority can be held responsible for them. PV and AM acknowledge support from the National Research Foundation of South Africa, and AM financial support from the Swedish International Development Cooperation Agency (SIDA) through the International Science Programme (ISP) -- Uppsala University to the University of Rwanda through the Rwanda Astrophysics, Space and Climate Science Research Group (RASCSRG). 





\bibliographystyle{mnras}
\bibliography{refs} 

\begin{thebibliography}{}
\makeatletter
\relax
\def\mn@urlcharsother{\let\do\@makeother \do\$\do\&\do\#\do\^\do\_\do\%\do\~}
\def\mn@doi{\begingroup\mn@urlcharsother \@ifnextchar [ {\mn@doi@} {\mn@doi@[]}}
\def\mn@doi@[#1]#2{\def\@tempa{#1}\ifx\@tempa\@empty \href {http://dx.doi.org/#2} {doi:#2}\else \href {http://dx.doi.org/#2} {#1}\fi \endgroup}
\def\mn@eprint#1#2{\mn@eprint@#1:#2::\@nil}
\def\mn@eprint@arXiv#1{\href {http://arxiv.org/abs/#1} {{\tt arXiv:#1}}}
\def\mn@eprint@dblp#1{\href {http://dblp.uni-trier.de/rec/bibtex/#1.xml} {dblp:#1}}
\def\mn@eprint@#1:#2:#3:#4\@nil{\def\@tempa {#1}\def\@tempb {#2}\def\@tempc {#3}\ifx \@tempc \@empty \let \@tempc \@tempb \let \@tempb \@tempa \fi \ifx \@tempb \@empty \def\@tempb {arXiv}\fi \@ifundefined {mn@eprint@\@tempb}{\@tempb:\@tempc}{\expandafter \expandafter \csname mn@eprint@\@tempb\endcsname \expandafter{\@tempc}}}

\bibitem[\protect\citeauthoryear{{Abbott} et~al.,}{{Abbott} et~al.}{2017}]{Abbott2017b}
{Abbott} B.~P.,  et~al., 2017, \mn@doi [\apjl] {10.3847/2041-8213/aa920c}, \href {https://ui.adsabs.harvard.edu/abs/2017ApJ...848L..13A} {848, L13}

\bibitem[\protect\citeauthoryear{{Andreoni} et~al.,}{{Andreoni} et~al.}{2017}]{Andreoni2017}
{Andreoni} I.,  et~al., 2017, \mn@doi [\pasa] {10.1017/pasa.2017.65}, \href {https://ui.adsabs.harvard.edu/abs/2017PASA...34...69A} {34, e069}

\bibitem[\protect\citeauthoryear{{Bauswein}, {Goriely}  \& {Janka}}{{Bauswein} et~al.}{2013}]{Bauswein2013}
{Bauswein} A.,  {Goriely} S.,   {Janka} H.~T.,  2013, \mn@doi [\apj] {10.1088/0004-637X/773/1/78}, \href {https://ui.adsabs.harvard.edu/abs/2013ApJ...773...78B} {773, 78}

\bibitem[\protect\citeauthoryear{{Buckley} et~al.,}{{Buckley} et~al.}{2018}]{Buckley2018}
{Buckley} D. A.~H.,  et~al., 2018, \mn@doi [\mnras] {10.1093/mnrasl/slx196}, \href {https://ui.adsabs.harvard.edu/abs/2018MNRAS.474L..71B} {474, L71}

\bibitem[\protect\citeauthoryear{{Burgh}, {Nordsieck}, {Kobulnicky}, {Williams}, {O'Donoghue}, {Smith}  \& {Percival}}{{Burgh} et~al.}{2003}]{Burgh2003}
{Burgh} E.~B.,  {Nordsieck} K.~H.,  {Kobulnicky} H.~A.,  {Williams} T.~B.,  {O'Donoghue} D.,  {Smith} M.~P.,   {Percival} J.~W.,  2003, in {Iye} M.,  {Moorwood} A. F.~M.,  eds,  Society of Photo-Optical Instrumentation Engineers (SPIE) Conference Series Vol. 4841, Instrument Design and Performance for Optical/Infrared Ground-based Telescopes. pp 1463--1471, \mn@doi{10.1117/12.460312}

\bibitem[\protect\citeauthoryear{{Chornock} et~al.,}{{Chornock} et~al.}{2017}]{Chornock2017}
{Chornock} R.,  et~al., 2017, \mn@doi [\apjl] {10.3847/2041-8213/aa905c}, \href {https://ui.adsabs.harvard.edu/abs/2017ApJ...848L..19C} {848, L19}

\bibitem[\protect\citeauthoryear{{Collins}, {Bauswein}, {Sim}, {Vijayan}, {Mart{\'\i}nez-Pinedo}, {Just}, {Shingles}  \& {Kromer}}{{Collins} et~al.}{2023}]{Collins2023}
{Collins} C.~E.,  {Bauswein} A.,  {Sim} S.~A.,  {Vijayan} V.,  {Mart{\'\i}nez-Pinedo} G.,  {Just} O.,  {Shingles} L.~J.,   {Kromer} M.,  2023, \mn@doi [\mnras] {10.1093/mnras/stad606}, \href {https://ui.adsabs.harvard.edu/abs/2023MNRAS.521.1858C} {521, 1858}

\bibitem[\protect\citeauthoryear{{Collins} et~al.,}{{Collins} et~al.}{2024}]{Collins2024}
{Collins} C.~E.,  et~al., 2024, \mn@doi [\mnras] {10.1093/mnras/stae571}, \href {https://ui.adsabs.harvard.edu/abs/2024MNRAS.tmp..614C} {}

\bibitem[\protect\citeauthoryear{{Coulter} et~al.,}{{Coulter} et~al.}{2017}]{Coulter2017}
{Coulter} D.~A.,  et~al., 2017, \mn@doi [Science] {10.1126/science.aap9811}, \href {https://ui.adsabs.harvard.edu/abs/2017Sci...358.1556C} {358, 1556}

\bibitem[\protect\citeauthoryear{{Domoto}, {Tanaka}, {Wanajo}  \& {Kawaguchi}}{{Domoto} et~al.}{2021}]{Domoto2021}
{Domoto} N.,  {Tanaka} M.,  {Wanajo} S.,   {Kawaguchi} K.,  2021, \mn@doi [\apj] {10.3847/1538-4357/abf358}, \href {https://ui.adsabs.harvard.edu/abs/2021ApJ...913...26D} {913, 26}

\bibitem[\protect\citeauthoryear{{Domoto}, {Tanaka}, {Kato}, {Kawaguchi}, {Hotokezaka}  \& {Wanajo}}{{Domoto} et~al.}{2022}]{Domoto2022}
{Domoto} N.,  {Tanaka} M.,  {Kato} D.,  {Kawaguchi} K.,  {Hotokezaka} K.,   {Wanajo} S.,  2022, \mn@doi [\apj] {10.3847/1538-4357/ac8c36}, \href {https://ui.adsabs.harvard.edu/abs/2022ApJ...939....8D} {939, 8}

\bibitem[\protect\citeauthoryear{{Drout} et~al.,}{{Drout} et~al.}{2017}]{Drout2017}
{Drout} M.~R.,  et~al., 2017, \mn@doi [Science] {10.1126/science.aaq0049}, \href {https://ui.adsabs.harvard.edu/abs/2017Sci...358.1570D} {358, 1570}

\bibitem[\protect\citeauthoryear{{Evans} et~al.,}{{Evans} et~al.}{2017}]{Evans2017}
{Evans} P.~A.,  et~al., 2017, \mn@doi [Science] {10.1126/science.aap9580}, \href {https://ui.adsabs.harvard.edu/abs/2017Sci...358.1565E} {358, 1565}

\bibitem[\protect\citeauthoryear{{Fitzpatrick}}{{Fitzpatrick}}{2004}]{Fitzpatrick2004}
{Fitzpatrick} E.~L.,  2004, in {Witt} A.~N.,  {Clayton} G.~C.,   {Draine} B.~T.,  eds,  Astronomical Society of the Pacific Conference Series Vol. 309, Astrophysics of Dust. p.~33 (\mn@eprint {arXiv} {astro-ph/0401344}), \mn@doi{10.48550/arXiv.astro-ph/0401344}

\bibitem[\protect\citeauthoryear{{Gillanders}, {Smartt}, {Sim}, {Bauswein}  \& {Goriely}}{{Gillanders} et~al.}{2022}]{Gillanders2022}
{Gillanders} J.~H.,  {Smartt} S.~J.,  {Sim} S.~A.,  {Bauswein} A.,   {Goriely} S.,  2022, \mn@doi [\mnras] {10.1093/mnras/stac1258}, \href {https://ui.adsabs.harvard.edu/abs/2022MNRAS.515..631G} {515, 631}

\bibitem[\protect\citeauthoryear{{Gillanders}, {Sim}, {Smartt}, {Goriely}  \& {Bauswein}}{{Gillanders} et~al.}{2023}]{Gillanders2023}
{Gillanders} J.~H.,  {Sim} S.~A.,  {Smartt} S.~J.,  {Goriely} S.,   {Bauswein} A.,  2023, \mn@doi [\mnras] {10.1093/mnras/stad3688}, \href {https://ui.adsabs.harvard.edu/abs/2023MNRAS.tmp.3536G} {}

\bibitem[\protect\citeauthoryear{{Green} et~al.,}{{Green} et~al.}{2018}]{Green2018}
{Green} G.~M.,  et~al., 2018, \mn@doi [\mnras] {10.1093/mnras/sty1008}, \href {https://ui.adsabs.harvard.edu/abs/2018MNRAS.478..651G} {478, 651}

\bibitem[\protect\citeauthoryear{{Hjorth} et~al.,}{{Hjorth} et~al.}{2017}]{Hjorth2017}
{Hjorth} J.,  et~al., 2017, \mn@doi [\apjl] {10.3847/2041-8213/aa9110}, \href {https://ui.adsabs.harvard.edu/abs/2017ApJ...848L..31H} {848, L31}

\bibitem[\protect\citeauthoryear{{Hotokezaka}, {Tanaka}, {Kato}  \& {Gaigalas}}{{Hotokezaka} et~al.}{2023}]{Hotokezaka2023}
{Hotokezaka} K.,  {Tanaka} M.,  {Kato} D.,   {Gaigalas} G.,  2023, \mn@doi [\mnras] {10.1093/mnrasl/slad128}, \href {https://ui.adsabs.harvard.edu/abs/2023MNRAS.526L.155H} {526, L155}

\bibitem[\protect\citeauthoryear{{Howlett} \& {Davis}}{{Howlett} \& {Davis}}{2020}]{Howlett2020}
{Howlett} C.,  {Davis} T.~M.,  2020, \mn@doi [\mnras] {10.1093/mnras/staa049}, \href {https://ui.adsabs.harvard.edu/abs/2020MNRAS.492.3803H} {492, 3803}

\bibitem[\protect\citeauthoryear{{Just} et~al.,}{{Just} et~al.}{2023}]{Just2023}
{Just} O.,  et~al., 2023, \mn@doi [\apjl] {10.48550/arXiv.2302.10928}, \href {https://ui.adsabs.harvard.edu/abs/2023ApJ...951L..12J} {951, L12}

\bibitem[\protect\citeauthoryear{{Kasen}, {Fern{\'a}ndez}  \& {Metzger}}{{Kasen} et~al.}{2015}]{Kasen2015}
{Kasen} D.,  {Fern{\'a}ndez} R.,   {Metzger} B.~D.,  2015, \mn@doi [\mnras] {10.1093/mnras/stv721}, \href {https://ui.adsabs.harvard.edu/abs/2015MNRAS.450.1777K} {450, 1777}

\bibitem[\protect\citeauthoryear{{Levan} et~al.,}{{Levan} et~al.}{2024}]{Levan2024}
{Levan} A.~J.,  et~al., 2024, \mn@doi [\nat] {10.1038/s41586-023-06759-1}, \href {https://ui.adsabs.harvard.edu/abs/2024Natur.626..737L} {626, 737}

\bibitem[\protect\citeauthoryear{{McCully} et~al.,}{{McCully} et~al.}{2017}]{McCully2017}
{McCully} C.,  et~al., 2017, \mn@doi [\apjl] {10.3847/2041-8213/aa9111}, \href {https://ui.adsabs.harvard.edu/abs/2017ApJ...848L..32M} {848, L32}

\bibitem[\protect\citeauthoryear{{Mooley} et~al.,}{{Mooley} et~al.}{2018}]{Mooley2018}
{Mooley} K.~P.,  et~al., 2018, \mn@doi [\nat] {10.1038/s41586-018-0486-3}, \href {https://ui.adsabs.harvard.edu/abs/2018Natur.561..355M} {561, 355}

\bibitem[\protect\citeauthoryear{{Mooley}, {Anderson}  \& {Lu}}{{Mooley} et~al.}{2022}]{Mooley2022}
{Mooley} K.~P.,  {Anderson} J.,   {Lu} W.,  2022, \mn@doi [Nature] {10.1038/s41586-022-05145-7}, \href {https://ui.adsabs.harvard.edu/abs/2022Natur.610..273M} {610, 273}

\bibitem[\protect\citeauthoryear{{Nicholl} et~al.,}{{Nicholl} et~al.}{2017}]{Nicholl2017}
{Nicholl} M.,  et~al., 2017, \mn@doi [\apjl] {10.3847/2041-8213/aa9029}, \href {https://ui.adsabs.harvard.edu/abs/2017ApJ...848L..18N} {848, L18}

\bibitem[\protect\citeauthoryear{{Nicolaou}, {Lahav}, {Lemos}, {Hartley}  \& {Braden}}{{Nicolaou} et~al.}{2020}]{Nicolaou2020}
{Nicolaou} C.,  {Lahav} O.,  {Lemos} P.,  {Hartley} W.,   {Braden} J.,  2020, \mn@doi [\mnras] {10.1093/mnras/staa1120}, \href {https://ui.adsabs.harvard.edu/abs/2020MNRAS.495...90N} {495, 90}

\bibitem[\protect\citeauthoryear{{Perego} et~al.,}{{Perego} et~al.}{2022}]{Perego2022}
{Perego} A.,  et~al., 2022, \mn@doi [\apj] {10.3847/1538-4357/ac3751}, \href {https://ui.adsabs.harvard.edu/abs/2022ApJ...925...22P} {925, 22}

\bibitem[\protect\citeauthoryear{{Pian} et~al.,}{{Pian} et~al.}{2017}]{Pian2017}
{Pian} E.,  et~al., 2017, \mn@doi [\nat] {10.1038/nature24298}, \href {https://ui.adsabs.harvard.edu/abs/2017Natur.551...67P} {551, 67}

\bibitem[\protect\citeauthoryear{{Planck Collaboration} et~al.,}{{Planck Collaboration} et~al.}{2016}]{Planck2016}
{Planck Collaboration} et~al., 2016, \mn@doi [\aap] {10.1051/0004-6361/201629022}, \href {https://ui.adsabs.harvard.edu/abs/2016A&A...596A.109P} {596, A109}

\bibitem[\protect\citeauthoryear{{Pognan}, {Jerkstrand}  \& {Grumer}}{{Pognan} et~al.}{2022}]{Pognan2022}
{Pognan} Q.,  {Jerkstrand} A.,   {Grumer} J.,  2022, \mn@doi [\mnras] {10.1093/mnras/stac1253}, \href {https://ui.adsabs.harvard.edu/abs/2022MNRAS.513.5174P} {513, 5174}

\bibitem[\protect\citeauthoryear{{Pognan}, {Grumer}, {Jerkstrand}  \& {Wanajo}}{{Pognan} et~al.}{2023}]{Pognan2023}
{Pognan} Q.,  {Grumer} J.,  {Jerkstrand} A.,   {Wanajo} S.,  2023, \mn@doi [\mnras] {10.1093/mnras/stad3106}, \href {https://ui.adsabs.harvard.edu/abs/2023MNRAS.526.5220P} {526, 5220}

\bibitem[\protect\citeauthoryear{{Shappee} et~al.,}{{Shappee} et~al.}{2017}]{Shappee2017}
{Shappee} B.~J.,  et~al., 2017, \mn@doi [Science] {10.1126/science.aaq0186}, \href {https://ui.adsabs.harvard.edu/abs/2017Sci...358.1574S} {358, 1574}

\bibitem[\protect\citeauthoryear{{Shingles} et~al.,}{{Shingles} et~al.}{2023}]{Shingles2023}
{Shingles} L.~J.,  et~al., 2023, \mn@doi [\apjl] {10.3847/2041-8213/acf29a}, \href {https://ui.adsabs.harvard.edu/abs/2023ApJ...954L..41S} {954, L41}

\bibitem[\protect\citeauthoryear{{Smartt} et~al.,}{{Smartt} et~al.}{2017}]{Smartt2017}
{Smartt} S.~J.,  et~al., 2017, \mn@doi [\nat] {10.1038/nature24303}, \href {https://ui.adsabs.harvard.edu/abs/2017Natur.551...75S} {551, 75}

\bibitem[\protect\citeauthoryear{{Sneppen}}{{Sneppen}}{2023}]{Sneppen2023_bb}
{Sneppen} A.,  2023, \mn@doi [\apj] {10.3847/1538-4357/acf200}, \href {https://ui.adsabs.harvard.edu/abs/2023ApJ...955...44S} {955, 44}

\bibitem[\protect\citeauthoryear{{Sneppen} \& {Watson}}{{Sneppen} \& {Watson}}{2023}]{Sneppen2023b}
{Sneppen} A.,  {Watson} D.,  2023, \mn@doi [\aap] {10.1051/0004-6361/202346421}, \href {https://ui.adsabs.harvard.edu/abs/2023A&A...675A.194S} {675, A194}

\bibitem[\protect\citeauthoryear{{Sneppen}, {Watson}, {Gillanders}  \& {Heintz}}{{Sneppen} et~al.}{2023a}]{Sneppen2023_rapid}
{Sneppen} A.,  {Watson} D.,  {Gillanders} J.~H.,   {Heintz} K.~E.,  2023a, \mn@doi [arXiv e-prints] {10.48550/arXiv.2312.02258}, \href {https://ui.adsabs.harvard.edu/abs/2023arXiv231202258S} {p. arXiv:2312.02258}

\bibitem[\protect\citeauthoryear{{Sneppen}, {Watson}, {Bauswein}, {Just}, {Kotak}, {Nakar}, {Poznanski}  \& {Sim}}{{Sneppen} et~al.}{2023b}]{Sneppen2023}
{Sneppen} A.,  {Watson} D.,  {Bauswein} A.,  {Just} O.,  {Kotak} R.,  {Nakar} E.,  {Poznanski} D.,   {Sim} S.,  2023b, \mn@doi [\nat] {10.1038/s41586-022-05616-x}, \href {https://ui.adsabs.harvard.edu/abs/2023Natur.614..436S} {614, 436}

\bibitem[\protect\citeauthoryear{{Sneppen}, {Watson}, {Poznanski}, {Just}, {Bauswein}  \& {Wojtak}}{{Sneppen} et~al.}{2023c}]{Sneppen2023A&A}
{Sneppen} A.,  {Watson} D.,  {Poznanski} D.,  {Just} O.,  {Bauswein} A.,   {Wojtak} R.,  2023c, \mn@doi [\aap] {10.1051/0004-6361/202346306}, \href {https://ui.adsabs.harvard.edu/abs/2023A&A...678A..14S} {678, A14}

\bibitem[\protect\citeauthoryear{{Tak}, {Uhm}  \& {Gillanders}}{{Tak} et~al.}{2024}]{Tak2024}
{Tak} D.,  {Uhm} Z.~L.,   {Gillanders} J.~H.,  2024, \mn@doi [arXiv e-prints] {10.48550/arXiv.2402.05471}, \href {https://ui.adsabs.harvard.edu/abs/2024arXiv240205471T} {p. arXiv:2402.05471}

\bibitem[\protect\citeauthoryear{{Tanaka} et~al.,}{{Tanaka} et~al.}{2023}]{Tanaka2023}
{Tanaka} M.,  et~al., 2023, \mn@doi [\apj] {10.3847/1538-4357/acdc95}, \href {https://ui.adsabs.harvard.edu/abs/2023ApJ...953...17T} {953, 17}

\bibitem[\protect\citeauthoryear{{Tanvir} et~al.,}{{Tanvir} et~al.}{2017}]{Tanvir2017}
{Tanvir} N.~R.,  et~al., 2017, \mn@doi [\apjl] {10.3847/2041-8213/aa90b6}, \href {https://ui.adsabs.harvard.edu/abs/2017ApJ...848L..27T} {848, L27}

\bibitem[\protect\citeauthoryear{{Tarumi}, {Hotokezaka}, {Domoto}  \& {Tanaka}}{{Tarumi} et~al.}{2023}]{Tarumi2023}
{Tarumi} Y.,  {Hotokezaka} K.,  {Domoto} N.,   {Tanaka} M.,  2023, \mn@doi [arXiv e-prints] {10.48550/arXiv.2302.13061}, \href {https://ui.adsabs.harvard.edu/abs/2023arXiv230213061T} {p. arXiv:2302.13061}

\bibitem[\protect\citeauthoryear{{Valenti} et~al.,}{{Valenti} et~al.}{2017}]{Valenti2017}
{Valenti} S.,  et~al., 2017, \mn@doi [\apjl] {10.3847/2041-8213/aa8edf}, \href {https://ui.adsabs.harvard.edu/abs/2017ApJ...848L..24V} {848, L24}

\bibitem[\protect\citeauthoryear{{Vieira}, {Ruan}, {Haggard}, {Ford}, {Drout}  \& {Fern{\'a}ndez}}{{Vieira} et~al.}{2023a}]{Vieira2023arxiv}
{Vieira} N.,  {Ruan} J.~J.,  {Haggard} D.,  {Ford} N.~M.,  {Drout} M.~R.,   {Fern{\'a}ndez} R.,  2023a, \mn@doi [arXiv e-prints] {10.48550/arXiv.2308.16796}, \href {https://ui.adsabs.harvard.edu/abs/2023arXiv230816796V} {p. arXiv:2308.16796}

\bibitem[\protect\citeauthoryear{{Vieira}, {Ruan}, {Haggard}, {Ford}, {Drout}, {Fern{\'a}ndez}  \& {Badnell}}{{Vieira} et~al.}{2023b}]{Vieira2023}
{Vieira} N.,  {Ruan} J.~J.,  {Haggard} D.,  {Ford} N.,  {Drout} M.~R.,  {Fern{\'a}ndez} R.,   {Badnell} N.~R.,  2023b, \mn@doi [\apj] {10.3847/1538-4357/acae72}, \href {https://ui.adsabs.harvard.edu/abs/2023ApJ...944..123V} {944, 123}

\bibitem[\protect\citeauthoryear{{Villar} et~al.,}{{Villar} et~al.}{2017}]{Villar2017}
{Villar} V.~A.,  et~al., 2017, \mn@doi [\apjl] {10.3847/2041-8213/aa9c84}, \href {https://ui.adsabs.harvard.edu/abs/2017ApJ...851L..21V} {851, L21}

\bibitem[\protect\citeauthoryear{{Watson} et~al.,}{{Watson} et~al.}{2019}]{Watson2019}
{Watson} D.,  et~al., 2019, \mn@doi [\nat] {10.1038/s41586-019-1676-3}, \href {https://ui.adsabs.harvard.edu/abs/2019Natur.574..497W} {574, 497}

\bibitem[\protect\citeauthoryear{{Waxman}, {Ofek}, {Kushnir}  \& {Gal-Yam}}{{Waxman} et~al.}{2018}]{Waxman2018}
{Waxman} E.,  {Ofek} E.~O.,  {Kushnir} D.,   {Gal-Yam} A.,  2018, \mn@doi [\mnras] {10.1093/mnras/sty2441}, \href {https://ui.adsabs.harvard.edu/abs/2018MNRAS.481.3423W} {481, 3423}

\bibitem[\protect\citeauthoryear{{Yang} et~al.,}{{Yang} et~al.}{2024}]{Yang2024}
{Yang} Y.-H.,  et~al., 2024, \mn@doi [\nat] {10.1038/s41586-023-06979-5}, \href {https://ui.adsabs.harvard.edu/abs/2024Natur.626..742Y} {626, 742}

\makeatother
\end{thebibliography}

\setcounter{section}{1}
\setcounter{equation}{0}
\setcounter{figure}{0}
\renewcommand{\thesection}{Appendix \arabic{section}}
\renewcommand{\theequation}{A.\arabic{equation}}
\renewcommand{\thefigure}{A.\arabic{figure}}

\section*{Appendix} 

\begin{figure}
    \centering
    \includegraphics[width=\linewidth,viewport=15 20 450 430 ,clip=]{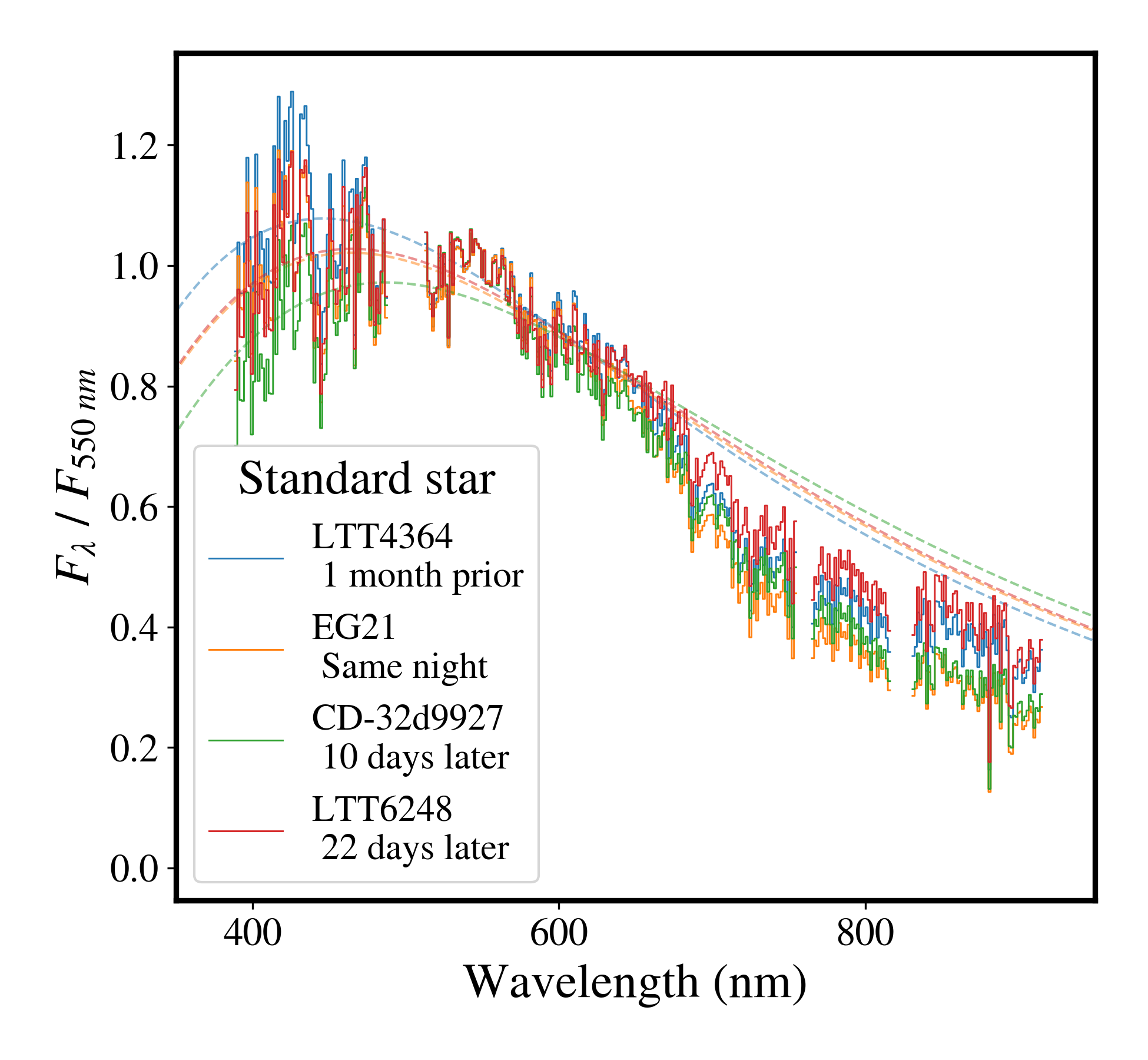}
    \caption{The spectral shape of the SALT spectrum at 1.17\,days using various standard stars for flux calibration. While the overall spectral shape is only weakly affected by the choice of standard star, the inferred depth of the absorption feature is smaller using spectrophotometric standards more than 1--2 weeks from the date of observation. } 
    \label{fig:SALT_standardstars}
\end{figure}

\subsection{ SALT Reduction }\label{sec:SALT_red}
As reported in \citet{Buckley2018}, the SALT spectrum was observed in the evening twilight of August 18, 2017, with the Robert Stobie Spectrograph \citep{Burgh2003} in long-slit mode, using a 2\arcsec\ slit-width with a low-resolution PG300 grating giving access to the whole visible wavelength range, with a mean $R\approx380$ spectral resolution. For this work, we used the originally-reduced data and 1D extraction, and used a wavelength range of approximately 380 to 900\,nm.

However, given the criticality of the overall spectral {\em shape} in this work, we revisited the flux calibration.  First of all, we did this independently, and several times, using different parameter choices for the functions and orders, when fitting the spectrophotometric standard star, EG21, observed during the same night. Moreover, to check the robustness of the results, we also used different spectrophotometric standards, observed with exactly the same instrumental configuration as searched and found from the SALT database.  We tested the calibration results using ones found closest to the date of observation: CD-32d9927, observed on 28~August, LTT4364 from 8~July, and LTT6248 from 10~September.  Our tests showed that all the various independent choices using the original and most appropriate star, EG21, result in the same overall shape, with only very minor differences irrelevant to the final results.  As a further demonstration of confidence, while the calibrations done with CD-32d992, LTT4364, and LTT6248 do show some variations in spectral shape\footnote{This is not unexpected due to the time difference; we note that normally SALT Operations do not consider using spectrophotometric standards taken more than 5--7\,days from the date of observations}, corresponding to 3\%-level differences in the best-fit temperature, the \srii absorption break location is similar in all cases beginning abruptly around the same wavelength, $\lambda \sim 640-680$\,nm.  Thus, even the variation between the standard stars does not impact the break, and we conclude that the shape of the SALT spectrum is fully robust for the analysis undertaken in this work.  The most appropriate spectrum, that calibrated with EG21, is displayed in all the figures and fits (see e.g.\ Fig~\ref{fig:sr_emergence}). 

Finally, we note that using calibrations with LTT4364 and LTT6248, i.e.\ the ones furthest away from the date of observation, the {\em depth} of the absorption feature will become somewhat smaller, translating to a difference in the feature's optical depth and by extension in the abundance required to fit the feature (see Fig.~\ref{fig:SALT_standardstars}). However, this is of minor concern to our analysis, as exact mass-estimates already have sizeable systematic uncertainties and as we (for testing line-identifications and constraining the ionisation-transition) are mainly concerned with whether the feature has emerged at this time or not.

\end{document}